\documentclass[utf8,10pt]{formatLJ}

\usepackage{url,lineno,microtype,subcaption}
\usepackage[onehalfspacing]{setspace}
\usepackage{bm}
\usepackage[numbers]{natbib}
\def\keyFont{\fontsize{8}{11}\helveticabold }
\def\firstAuthorLast{Jeancolas {et~al.}} 
\def\Authors{Laetitia Jeancolas\,$^{1,2*}$, Dijana Petrovska-Delacrétaz\,$^{2}$, Graziella Mangone\,$^{3,4}$,\\
 Badr-Eddine Benkelfat\,$^{2}$, Jean-Christophe Corvol\,$^{3,4}$, Marie Vidailhet\,$^{3,4}$,\\
  Stéphane Lehéricy\,$^{1,3,5}$ and Habib Benali\,$^{6}$}
\def\Address{$^{1}$ Paris Brain Institute - ICM, Centre de NeuroImagerie de Recherche – CENIR, Paris, France\\
$^{2}$ SAMOVAR, Télécom SudParis, Institut Polytechnique de Paris, Evry, France\\
$^{3}$ Sorbonne University, Inserm, CNRS, Paris Brain Institute - ICM, Paris, France\\
$^{4}$ Assistance Publique Hôpitaux de Paris, Hôpital Pitié-Salpêtrière, Department of Neurology, \\
Clinical Investigation Center for Neurosciences, Paris, France \\
$^{5}$ Assistance Publique Hôpitaux de Paris, Hôpital Pitié-Salpêtrière, Department of Neuroradiology, Paris, France\\
$^{6}$ PERFORM Center,  Concordia University, Electrical \& Computer Engineering Department, Montreal, Canada}

\def\corrAuthor{Laetitia Jeancolas}

\def\corrEmail{\href{mailto:laetitia.jeancolas@icm-institute.org}{laetitia.jeancolas@icm-institute.org}}

\begin{document}

\pdfoutput=1
\raggedbottom
\firstpage{1}

\title[Parkinson's Detection from X-vectors]{X-vectors: New Quantitative Biomarkers for Early Parkinson's Disease Detection from Speech } 

\author[\firstAuthorLast ]{\Authors} 
\address{} 
\correspondance{} 

\extraAuth{}

\maketitle



\begin{abstract}

Many articles have used voice analysis to detect Parkinson's disease (PD), but few have focused on the early stages of the disease and the gender effect. In this article, we have adapted the latest speaker recognition system, called x-vectors, in order to detect an early stage of PD from voice analysis. X-vectors are embeddings extracted from a deep neural network, which provide robust speaker representations and improve speaker recognition when large amounts of training data are used.\\
Our goal was to assess whether, in the context of early PD detection, this technique would outperform the more standard classifier MFCC-GMM (Mel-Frequency Cepstral Coefficients - Gaussian Mixture Model) and, if so, under which conditions. \\
We recorded 221 French speakers (including recently diagnosed PD subjects and healthy controls) with a high-quality microphone and with their own telephone. Men and women were analyzed separately in order to have more precise models and to assess a possible gender effect. 
Several experimental and methodological aspects were tested in order to analyze their impacts on classification performance. We assessed the impact of audio segment duration, data augmentation, type of dataset used for the neural network training, kind of speech tasks, and back-end analyses.
X-vectors technique provided better classification performances than MFCC-GMM for text-independent tasks, and seemed to be particularly suited for the early detection of PD in women (7 to 15\% improvement). This result was observed for both recording types (high-quality microphone and telephone).

\keyFont{ \section{Keywords:} Parkinson's disease, x-vectors, voice analysis, MFCC, early detection, automatic detection, telediagnosis}

\end{abstract}

\section{Introduction}

Parkinson's disease (PD) is the second most common neurodegenerative disease after Alzheimer's disease and affects approximately seven million people worldwide. Its prevalence in industrialized countries is around 0.3\% and increases with age: 1\% of people over the age of 60 and up to 4\% of those over 80 are affected \cite{de_lau_epidemiology_2006}. The prevalence of PD has doubled between 1990 and 2016, which may be explained by the rise in life expectancy, better diagnoses and environmental factors.
 This disease results in motor disorders worsening over time caused by a progressive loss of dopaminergic neurons in the substantia nigra (located in the midbrain). The standard diagnosis is mainly based on clinical examination. Usually the diagnosis is made when at least two of the following three symptoms are noted: akinesia (slowness of initiation of movement), rigidity and tremors at rest. Unfortunately, these motor symptoms appear once 50 to 60\% of dopaminergic neurons in the substantia nigra \cite{haas_premotor_2012} and 60 to 80\% of their striatal endings \cite{fearnley_ageing_1991} have degenerated. That is why detecting PD in the early stages remains a big challenge, in order to test treatments before the occurrence of large irreversible brain damages, and later to slow down, or even stop, its progression from the beginning.

Voice impairment is one of the first symptoms to appear.  Many articles have used voice analysis to detect PD. They observed vocal disruptions, called hypokinetic dysarthria, expressed by a reduction in prosody, irregularities in phonation and difficulties in articulation. The classification performances (accuracy rate) using voice analysis ranged from 65 to 99\% for moderate to advanced stages of the disease. Fewer studies focused on early detection of PD through voice. Moreover, they usually worked on rather small databases (around 40 subjects) and analyzed men or mixed-gender groups
\cite{rusz_quantitative_2015, orozco-arroyave_automatic_2016, novotny_automatic_2014, rusz_acoustic_2011}. Recently, PD detection using telephone recordings has been carried out in the early stages \cite{jeancolas_comparison_2019} as well as in all stages combined \cite{arora_developing_2019}.

Different classification methodologies have been explored to detect PD from voice. 
The first studies used global features, such as the number of pauses, the number of dysfluent words, the standard deviation (SD) of pitch and of intensity, along with averaged low-level perturbations, such as shimmer, jitter, voice onset time, signal to noise ratio, formants or vowel space area, reviewed in \cite{jeancolas_analyse_2016}. The authors usually performed features selection, keeping statistically significant features and removing the redundancies. Finally, selected features were fed to classifiers, such as Support Vector Machines (SVM) \cite{little_suitability_2009, gil_diagnosing_2009, rusz_acoustic_2011, sakar_collection_2013, novotny_automatic_2014, rusz_speech_2015, sakar_analyzing_2017}, k-nearest neighbors \cite{sakar_collection_2013, sakar_analyzing_2017}, decision trees \cite{mucha_identification_2017}, multilayer perceptrons \cite{gil_diagnosing_2009}, probabilistic neural networks \cite{ene_neural_2008} or minimax classifiers with gaussian kernel density \cite{rusz_imprecise_2013}.

Other type of features has been used in the field of speaker recognition for decades: the Mel-Frequency Cepstral Coefficients (MFCC) \cite{bimbot_tutorial_2004}. These short-term features, calculated on [20-40ms] windows, characterize the spectral envelope and reflect the shape of the vocal tract. Over the past fifteen years, we have started to encounter them in the detection of vocal pathologies, such as dysphonia \cite{dibazar_feature_2002, godino-llorente_automatic_2004, malyska_automatic_2005}. The use of MFCC for PD detection was introduced in 2012 by Tsanas et al. \cite{tsanas_novel_2012}. Since then, many studies have used MFCC for PD detection, sometimes combining them with other features.

Several statistical analyses and classifiers can be applied on MFCC features. 
For instance, if MFCC dispersion is low within classes, generally due to poor phonetic variety, one can simply consider the MFCC averages (in addition to other features). This is generally the case for sustained vowel tasks \cite{tsanas_novel_2012, jafari_classification_2013, benba_voice_2014, benba_discriminating_2016, orozco-arroyave_characterization_2015, hemmerling_automatic_2016}  or when phonetically similar frames are selected \cite{orozco-arroyave_automatic_2014, orozco-arroyave_automatic_2016, orozco-arroyave_voiced/unvoiced_2015}. Authors often add to the means some other statistics like standard deviation, kurtosis (flattening measurement) and skewness (asymmetry measurement) in order to gain a little more information. These features are then fed into classifiers such as SVM, multilayer perceptrons or decision trees. 
 
If frames are acoustically very different (such as during whole reading or free speech tasks), additional precision is required to describe the MFCC distribution. One possible modeling is to use vector quantization \cite{benba_voice_2014, kapoor_parkinsons_2011}. Another more precise way is to model MFCC distribution with Gaussian Mixture Models (GMM). GMM can model MFCC distribution of PD and HC groups. Likelihood scores of test subject MFCC against the two GMM models (PD and control) are then calculated \cite{jeancolas_comparison_2019, moro-velazquez_analysis_2018}. GMM can also model the MFCC distribution of each subject. Means of Gaussian functions (forming a "supervector") are then fed into a classifier such as SVM \cite{bocklet_automatic_2013}. When not enough speech data is available to train the GMM models, which mainly occurs when GMM are used to model each subject (rather than a group), GMM can be adapted from Universal Background Models (UBM) previously trained with a bigger dataset \cite{reynolds_speaker_2000, bocklet_automatic_2013}. More than that, a more recent speaker recognition technique, called i-vectors, has been adapted for PD detection \cite{garcia_language_2017, moro-velazquez_analysis_2018}. This approach consists in removing the UBM mean supervector and projecting each supervector onto a low dimensional space, called total variability space. Intra-class variability is then often handled by means of discriminant techniques, like Linear Discriminant Analysis (LDA) or Probabilistic Linear Discriminant Analysis (PLDA). In PD detection this results in compensating the speaker, channel and session effects. In \cite{lopez_assessing_2019} the authors compared the i-vectors system with another MFCC-based speaker representation, using Fisher vectors, and found superior PD detection performance for the latter.

Over the last few years, with the increase of computing power, several Deep Neural Network (DNN) techniques have emerged in PD detection. Some studies applied Convolutional Neural Networks on spectrograms \cite{vasquez-correa_convolutional_2017, zhang_deepvoice:_2018, khojasteh_parkinsons_2018}. Others used DNN to extract phonological features from MFCC \cite{garcia-ospina_phonological_2018}, or to detect directly PD from global features \cite{rizvi_lstm_2020}.

 In the present study, we adapted a brand-new text-independent (i.e. no constraint on what the speaker can say) speaker recognition methodology, called x-vectors, introduced in 2016 \cite{snyder_deep_2016}. This approach consists in extracting embedding features from a DNN taking MFCC as inputs. According to the authors, classification from these features resulted in a more robust speaker representation \cite{snyder_deep_2017} and improved recognition, provided a large amount of training data \cite{snyder_x-vectors:_2018}. 
 In 2018, the same authors adapted the x-vector method to language recognition \cite{snyder_spoken_2018} and outperformed several state-of-the-art i-vector systems.
 
 Recently, we proposed an adaptation of x-vectors for PD detection in \cite{jeancolas_detection_2019}. Since then, another work has used x-vectors for PD detection \cite{moro-velazquez_using_2020}.
 In our paper we made different experimental choices. Unlike \cite{moro-velazquez_using_2020}, we focused on PD detection at an early stage, and performed the classifications on high-quality recordings on the one hand and on telephone recordings on the other hand. We also tested different types of speech tasks (text-dependent or text-independent) and different datasets for the DNN training, in order to assess their impact on PD detection.
 In order to achieve the best performance, we also considered men and women separately. This is usually done in speaker recognition and has been proved to enhance vocal pathology detections involving MFCC features \cite{fraile_automatic_2009}. Moreover, this allowed to analyze the effect of gender on PD detection from speech. 
 We also made different methodological choices. We studied the effect of important x-vectors methodological aspects, such as the audio segment duration and data augmentation. Finally we assessed the advantage of considering an ensemble method for the classification.
For each condition, we compared different classifiers: cosine distance (with and without LDA) and PLDA, which are commonly used with x-vectors, and as a baseline, the MFCC-GMM technique we used in \cite{jeancolas_comparison_2019}. 


\section{Materials and methods}

\subsection{Databases}

\subsubsection{Participants}

A total of 221 French speakers were included into this study: 121 PD patients and 100 healthy controls (HC). All PD patients and 49 HC were recruited at the Pitié-Salpêtrière Hospital, included in the ICEBERG cohort, a longitudinal observational study conducted at the Clinical Investigation Center for Neurosciences at the Paris Brain Institute (ICM). An additional 51 HC were recruited to balance the number of PD and control subjects. 
All patients had a diagnosis of PD according to UKPDSBB criteria with less than 4 years disease duration, and HC were free of any neurological disease or symptoms. All HC controls had a neurological examination to exclude subjects with parkinsonism or other neurological disease. Participants had neurological examination, motor and cognitive tests, biological sampling and brain MRI. 
 PD patients were pharmacologically treated and their voice were recorded during ON-state (less than 12 hours after their last medication intake).
Data from participants with technical recording issues, language disorders not related to PD (such as stuttering) or when deviation from the standardized procedure occurred, were excluded from the analysis.
The ICEBERG cohort (clinicaltrials.gov, NCT02305147) was conducted according to Good Clinical Practice guidelines. All participants received informed consent prior to any investigation. The study was sponsored by Inserm, and received approval from an ethical committee (IRBParis VI, RCB: 2014-A00725-42) according to local regulations.

\subsubsection{High-quality microphone recordings}
Among the 217 participants kept for the analysis, 206 subjects including 115 PD (74 males, 41 females) and 91 HC (48 males, 42 females) performed speech tasks recorded with a high-quality microphone. Information about age, duration since diagnosis, Hoehn \& Yahr stage \cite{hoehn_parkinsonism:_1967}, MDS-UPDRS III score \cite{goetz_movement_2007} (OFF state) and Levodopa Equivalent Daily Dose (LEDD) are detailed in Table \ref{table_database_pro}.
The microphone was a professional head mounted omnidirectional condenser microphone (Beyerdynamics Opus 55 mk ii) placed approximately 10 cm from the mouth. This microphone was connected to a professional sound card (Scarlett 2i2, Focusrite) which provided phantom power and pre-amplification. Speech was sampled at 96000 Hz with 24 bits resolution and a spectrum of [50Hz-20kHz]. ICEBERG participants were recorded in consultation rooms in the clinical investigation center and sleep disorder unit of the Pitié-Salpêtrière hospital in Paris. Additional HC were recorded in quiet rooms at their house or their office with the same recording devices.
Speech tasks were presented in a random order to the participants via a graphical user interface. Tasks which are analyzed in the present study are: readings (1min), sentence repetitions (10s), free speech (participants were asked to talk about their day during 1min)  and fast syllable repetitions (1min30), also called diadochokinesia (DDK) tasks (/pataka/, /badaga/, /pabikou/...).

\begin{table*}[tb]
\centering
\caption{High-quality microphone database information.\\
}
\begin{tabular}{ccccccc}
                   &   Number    & Age (years)  & Disease duration (years) & H \& Y & MDS-UPDRS III & LEDD (mg) \\
                   &           & mean $\pm$ SD & mean $\pm$ SD   & mean $\pm$ SD   & mean $\pm$ SD & mean $\pm$ SD\\
\hline
\rule[0cm]{0cm}{0.4cm}
\textbf{PD}&\textbf{115}& \textbf{63.8 $\pm$ 9.3}&\textbf{2.6 $\pm$ 1.5}&\textbf{2.0 $\pm$ 0.1}&\textbf{32.5  $\pm$ 7.0}&  \textbf{392 $\pm$ 266}\\
 M   &  74  &  63.7  $\pm$ 9.3   &  2.5 $\pm$ 1.4 &  2.0 $\pm$ 0.1  &  34.1  $\pm$ 7.0  &  415  $\pm$ 298 \\
  F   &  41  &  63.9  $\pm$ 9.3   &  2.7 $\pm$ 1.5 &  2.0 $\pm$ 0.0  &  29.6  $\pm$ 5.8  &   352  $\pm$ 191 \\
   \hline
   \rule[0cm]{0cm}{0.4cm}
\textbf{HC}&\textbf{91}&\textbf{59.1  $\pm$ 10.0}&   -    &\textbf{0.0 $\pm$ 0.3}&\textbf{4.8  $\pm$ 3.5}& - \\
       M   &  48  &  58.9  $\pm$ 10.7   &   -    &  0.0 $\pm$ 0.0  &  4.6  $\pm$ 3.7  & - \\
        F   &  43  &  59.3  $\pm$ 9.2   &   -    &  0.1 $\pm$ 0.4  &  4.9  $\pm$ 3.4  & - \\
   \hline
   \rule[0cm]{0cm}{0.4cm}
\textbf{Total  }&  \textbf{206}  &  \textbf{61.7  $\pm$ 9.8}   &    -   & \textbf{1.5 $\pm$ 0.9} &  \textbf{24.8  $\pm$ 13.9}  & - \\
\end{tabular}
\label{table_database_pro}
\end{table*}

\subsubsection{Telephone recordings}
Most of the participants, 101 PD (63 males, 38 females) and 61 HC (36 males, 25 females) also carried out telephone recordings at home. Information about age, duration since diagnosis, Hoehn \& Yahr stage, MDS-UPDRS III score (OFF state) and LEDD are detailed in Table \ref{table_database_tel}.
Participants called once a month with their own phone (mobile or landline) an interactive voicemail (IVM, from NCH company), connected to a SIP (Session Initiation Protocol) server (ippi). Audio signal was compressed with G711 codec and transformed into PCM16 audio files by IVM. Finally, speech files were sampled at 8000Hz with 16 bits resolution, and a frequency bandwidth of [300-3400Hz].
We set up the voicemail to automatically make the participants carry out a set of speech tasks when they called. Participants performed different numbers of recording sessions (from 1 to 13 with an average of 5) depending on when they started and early stoppings. Tasks that we analyzed in this study were: sentence repetitions (20s), free speech (1min) and DDK tasks (1min). Reading was not performed by telephone, because for practical reasons we wanted all the instructions to be audio. Details about the experimental setup (such as speech task content, transmission chain or encoding) were presented in \cite{jeancolas_detection_2019}.

\begin{table*}[tb]
\centering
\caption{Telephone database information. \\
}
\begin{tabular}{ccccccc}
                   &   Number    & Age (years)  & Disease duration (years) & H \& Y & MDS-UPDRS III & LEDD (mg) \\
                   &           & mean $\pm$ SD & mean $\pm$ SD   & mean $\pm$ SD   & mean $\pm$ SD & mean $\pm$ SD\\
       \hline
\rule[0cm]{0cm}{0.4cm}
\textbf{PD}&\textbf{101}& \textbf{63.5 $\pm$ 9.0}&\textbf{2.6 $\pm$ 1.4}&\textbf{2.0 $\pm$ 0.1}&\textbf{32.4  $\pm$ 7.0}  &  \textbf{387 $\pm$ 272}\\
 M   &  63  &  63.7  $\pm$ 9.0   &  2.5 $\pm$ 1.4 &  2.0 $\pm$ 0.1  &  34.2  $\pm$ 6.9  & 403  $\pm$ 311\\
  F   &  38  &  63.3  $\pm$ 9.3   &  2.7 $\pm$ 1.5 &  2.0 $\pm$ 0.0  &  29.5  $\pm$ 6.1  & 359  $\pm$ 194\\
   \hline
   \rule[0cm]{0cm}{0.4cm}
\textbf{HC}&\textbf{61}&\textbf{62.6  $\pm$ 8.5}&   -    &\textbf{0.0 $\pm$ 0.3}&\textbf{4.9  $\pm$ 3.5}& - \\
       M   &  36  &  63.1  $\pm$ 9.3   &   -    &  0.0 $\pm$ 0.0  &  4.6  $\pm$ 3.5  & - \\
        F   &  25  &  61.8  $\pm$ 7.4   &   -    &  0.1 $\pm$ 0.5  &  5.3  $\pm$ 3.6  & - \\
   \hline
   \rule[0cm]{0cm}{0.4cm}
\textbf{Total  }&  \textbf{162}  &  \textbf{63.2  $\pm$ 8.9}   &    -   & \textbf{1.4 $\pm$ 0.9} &  \textbf{23.9  $\pm$ 14.1}  & - \\
\end{tabular}
\label{table_database_tel}
\end{table*}

\subsection{Methods}

\subsubsection{Baseline: MFCC-GMM methodology}

In this section we present our MFCC-GMM baseline framework. This method, based on Gaussian mixture models fitting cepstral coefficients distribution of each class, has been used for decades in speaker recognition and was recently adapted for PD detection \cite{jeancolas_comparison_2019}.

\paragraph{Preprocessing and MFCC extraction}

The first preprocessing regarding our high-quality microphone recordings was spectral subtraction \cite{boll_suppression_1979}. The aim of this denoising technique was to compensate for the mismatched environments, by removing additive and stationary noises. We applied it with the Praat software \cite{boersma_praat_2001}, using the 5s silence recorded at the end of each participant's session for the calibration. Regarding the telephone recordings, spectral subtraction was not performed because acoustic environments were not different between PD subjects and HC.

We then extracted  the log-energy and 19 MFCC, using Kaldi software \cite{povey_kaldi_2011}, on 20ms overlapping windows, with a 10ms step. For the high-quality recordings, the 23 triangular mel bins covered a frequency range of  [20-7000Hz]. As for the telephone recordings, the frequency range of the mel bins was [300-3700Hz]. More details about the MFCC extraction methodology can be found in \cite{jeancolas_detection_2019}. 
The first derivatives (Deltas) and second derivatives (Delta-Deltas) were then computed and added to the feature vectors.

Once the MFCC and their deltas extracted, we carried out  Vocal Activity Detection (VAD), based on the log-energy, in order to remove silent frames. 

Finally, to complete denoising, cepstral mean subtraction \cite{quatieri_discrete-time_2001} was performed on 300ms sliding windows, reducing linear convolutional channel effects on both databases.

\paragraph{Distribution modeling with Gaussian Mixture Models}

We split the databases into three groups per gender: one group of PD subjects and one group of controls for training, and the remaining PD and control participants for testing. 
In the laboratory setting database, we took 36 PD and 36 HC for the male training groups and 38 PD and 12 HC for the male test group. As for women, we considered 30 PD and 30 HC for training and 11 PD and 13 HC for the test. 
For the telephone database, we selected 30 PD and 30 HC for the male training groups and 33 PD and 6 HC for the male test group. For females we used 20 PD and 20 HC for training and 18 PD and 5 HC for the test.

\begin{figure*}[htb]
      \centering
      \includegraphics[scale=0.7]{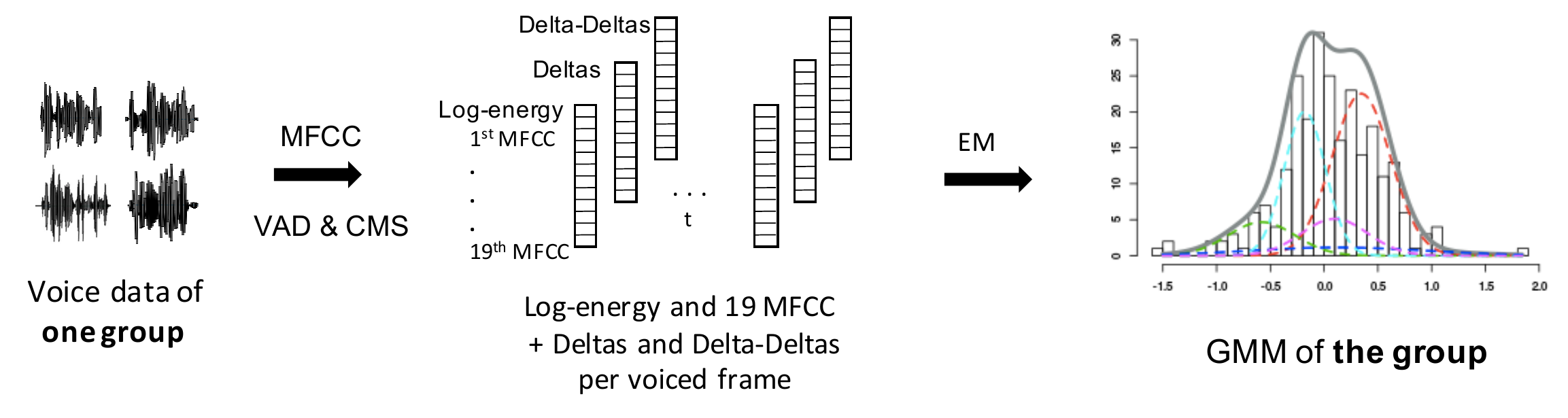}
     \caption{MFCC-GMM training phase: GMM training for each group (male PD,
female PD, male control and female control). VAD: Voice Activity Detection, CMS: Cepstral Mean Subtraction, EM: Expectation-Maximization}
     \label{schema_entrainement_GMM}
   \end{figure*}

During the training phase, we built multidimensional GMM to model the MFCC distributions of each training group (see Figure \ref{schema_entrainement_GMM}). Means, SD and weights of the Gaussians (characterizing the GMM) were estimated via an Expectation-Maximization algorithm.
The optimal number of Gaussian functions depends on quantity of speech data used for training. We chose 20 Gaussian functions for the present analyses on high-quality microphone database and 50 for the telephone database, as more sessions per subject were available.

\paragraph{Classification}

\begin{figure*}[htb]
      \centering
      \includegraphics[scale=0.55]{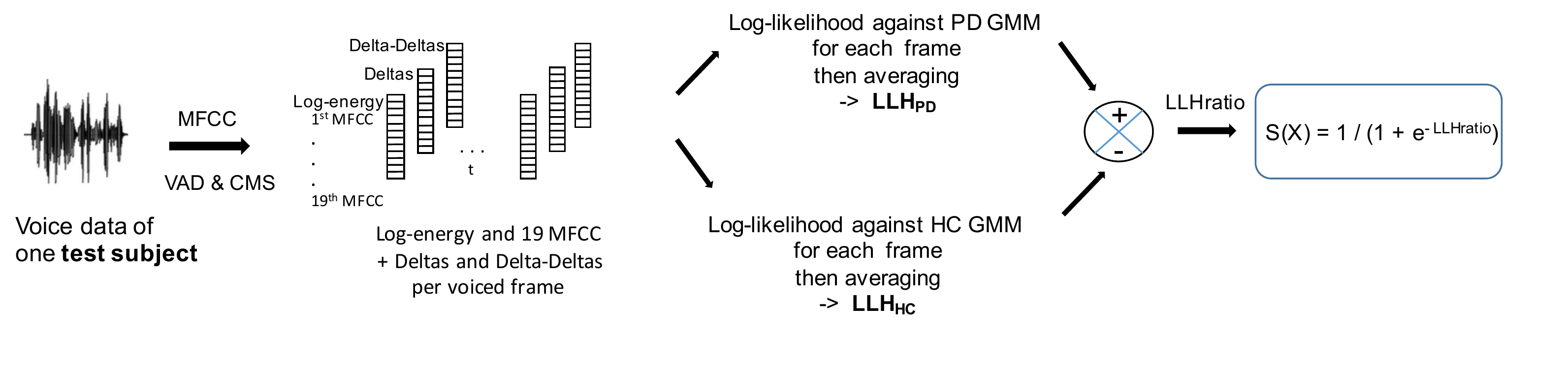}
     \caption{MFCC-GMM test phase: each test subject data are tested against a PD GMM model and a HC GMM model. The sigmoid of the log-likelihood ratio provides the classification score. VAD: Voice Activity Detection, CMS: Cepstral Mean Subtraction, LLH: log-likelihood.}
     \label{schema_test_GMM}
   \end{figure*}

For each test subject we calculated the log likelihood (LLH) of their MFCC compared to the two GMM models corresponding to their gender. We first computed one log-likelihood per frame (after silence removal) of the test subject data against the two models, then we took the average over all the frames. Thus, the likelihood was guaranteed to be independent of the number of frames.
 A sigmoid function was then applied to the difference of these means (the \textit{log-likelihood ratio}), so as to produce an S score ranging from 0 to 1 per test subject (see Figure \ref{schema_test_GMM}). A score close to 1 indicated a greater probability that the test subject had PD and a score close to 0 that he was healthy. 

\paragraph{Validation and ensemble method}
\label{section_agg_metho}
 
\begin{figure}[htb]
      \centering
      \includegraphics[scale=0.55]{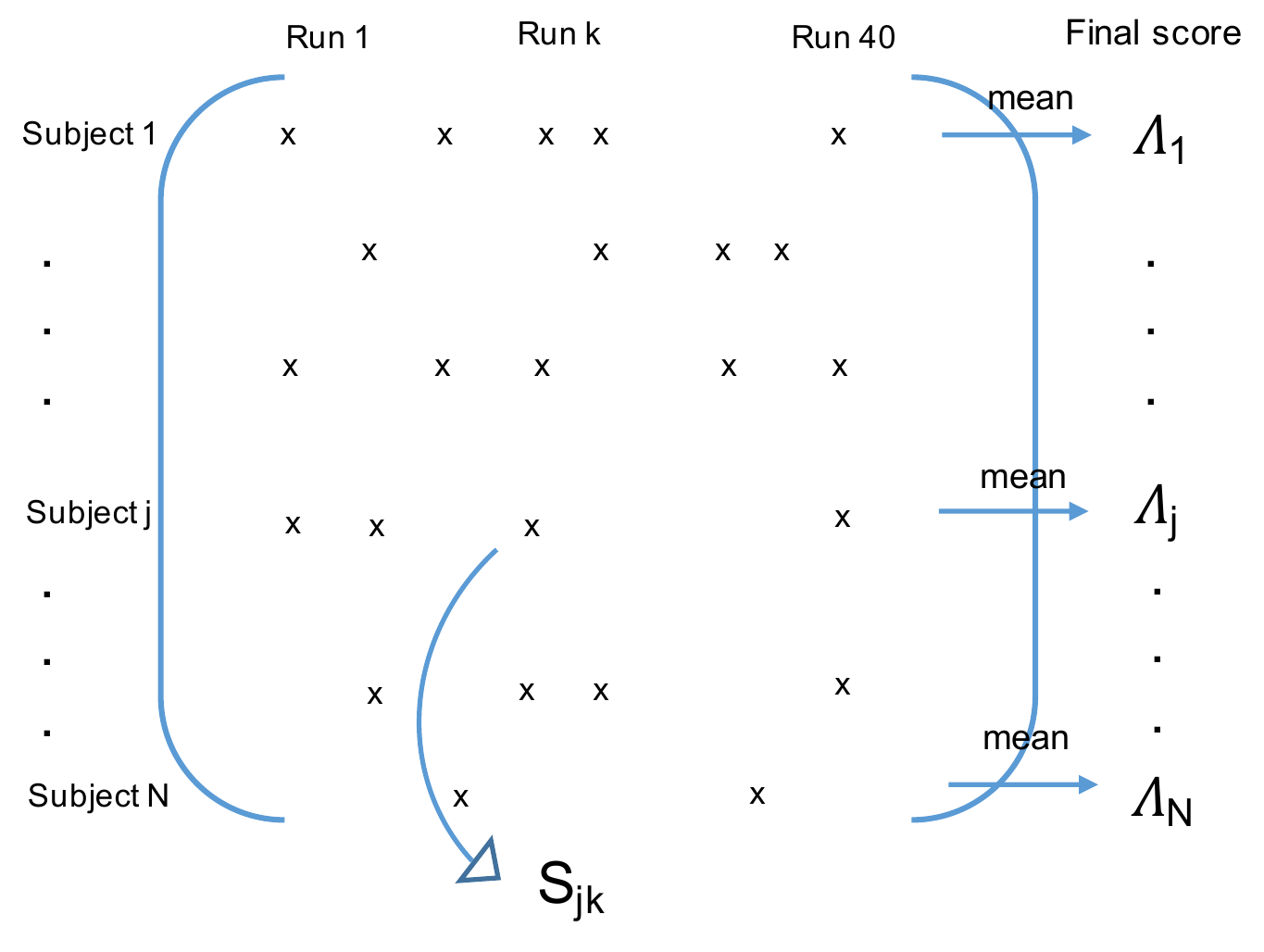}
     \caption{Ensemble method: repeated random subsampling  aggregation. Classification score $S_{jk}$ is the intermediate score of subject $j$ for run $k$. Final score $\Lambda_{j}$ is the average of the intermediate scores obtained during the 40 runs.} 
     \label{schema_validation}
   \end{figure}

The final classification was carried out with an ensemble method, a repeated random subsampling aggregation \cite{buhlmann_analyzing_2002, maillard_cross-validation_2017}, which is a type of bootstrap aggregation \cite{breiman_bagging_1996} without replacement.  
We ran 40 times GMM modeling and classification phases, each time with a different random split of participants between the training and test groups. Numbers of subjects per group were the ones previously stated. At the end of the 40 runs,  all the subjects were tested about ten times. For each subject, we finally averaged his classification scores obtained during the runs when he belonged to the test group (see Figure  \ref{schema_validation}). 

The choice of this ensemble method was based on several elements:

   \begin{itemize}
   \item First of all, regarding the sampling technique, we chose repeated random subsampling rather than k-fold or LOSO (which are more common) because it allowed us to have the same number of PD and HC subjects for training. This led to same training conditions for GMM, as same optimal number of Gaussians, therefore fewer hyperparameters and so a reduced risk of overfitting.
   
   \item We then chose to complete this cross-validation with an ensemble method, because they are known to decrease prediction variance, leading to usually better classification performance \cite{friedman_elements_2001}. 
       
   \item Regarding the type of aggregation, we chose to average the scores rather than using a majority vote type, because it is the technique which is known to minimize the variance the most \cite{friedman_elements_2001}.
   
   \item The error calculated on the final scores (of \textit {out-of-bag} type) is known to be a good unbiased estimate of the real (or generalized) error, namely the one we would have if we tested an infinity of other new subjects on our aggregated model.
   \end{itemize}

In section \ref{section_agg_result} we compared the classification performance of the aggregated model  with the performance of the simple model. The real performance of the simple model (the one we would have if we tested an infinity of other new subjects against two GMM trained with our current database) was estimated by the performance of the repeated random subsampling cross validation (i.e. the average of the classification performance of each run). In all other sections we used the aggregated model for the classification.

\subsubsection{X-vector methodology}

In this section we present the x-vector system we adapted from the latest speaker recognition method \cite{snyder_x-vectors:_2018}. 
X-vectors are fixed-length representations of variable-length speech segments. They are embeddings extracted from a feed-forward DNN taking MFCC vectors as input. Once extracted we classified the x-vectors with different classification methods (cosine distance, LDA + cosine distance and PLDA). 

\paragraph{DNN training}

Since DNN training usually requires a lot of data, we used DNN trained on large speaker recognition databases and available online (\url{http://kaldi-asr.org/models.html}).

For the analysis of our telephone recordings, we considered the pretrained DNN SRE16 model, described in \cite{snyder_x-vectors:_2018}. This DNN was trained on 5139 subjects from LDC catalog databases, including the Swichboard (Phase1,2,3 and Cellular 1,2), Mixer 6 and NIST SREs corpora. These databases contain telephone conversations and data recorded with a microphone, with English as the dominant language. Some data were directly sampled at 8 kHz, and the 16 kHz sampled recordings were then downsampled to 8 kHz. \\
For the analysis of our high-quality microphone recordings, we used the voxceleb model, trained on the voxceleb database \cite{nagrani_voxceleb:_2017}. Data came from video interviews of 7330 celebrities posted on Youtube. Voices were sampled at 16kHz. \\
Finally, data augmentation, as described in section \ref{section_aug_metho}, was applied to all these DNN training datasets. 

Those DNN were trained in the context of speaker identification. Inputs are log energy and MFCC extracted every 10ms from [2-4s] audio segments. For SRE16 model, 23 MFCC were extracted with a MEL bin range of [20-3700Hz]. For voxceleb model, 30 MFCC were extracted with a bin range of [20-7600Hz]. As for the MFCC-GMM analysis, a voice activity detection and cepstral mean subtraction were performed. Deltas and Delta-Deltas  were not computed because the temporal context was already taken into account in the DNN. 

DNN architecture is detailed in Table \ref{table_DNN}. The neural networks were composed of 3 parts:

- A set of frame-level layers taking MFCC as inputs. These layers constituted a Time Delay Neural Network (TDNN) taking into account a time context coming from neighboring frames.

- A statistics pooling layer aggregating the outputs (taking mean and SD) of the TDNN network across the audio segment. The output of this step was a large-scale (3000 dimensional) representation of the segment. 

- Last part was a simple feed forward network composed of two segment-level layers taking as input the result of the pooling layer, reducing its dimensionality to 512, and ending with a softmax layer. The softmax layer gave the probabilities that the input segment came from each speaker of the training database.

For the results presented in section \ref{section_ourDNN}, we trained a DNN  with our own data (telephone recordings). The only difference in the DNN architecture was the size of the softmax layer output, which was two. Indeed, here the DNN was trained directly to discriminate PD subjects from HC (two classes) instead of speakers (N classes).

\begin{table}[htb]
\centering
\caption{Embedding DNN architecture. X-vectors are extracted at layer segment-level 6 before Rectified Linear Unit (ReLU) activation function. T is the number of frames composing the input segment. K corresponds to the number of input features for one frame, K=24 for the telephone recordings (23 MFCC + log energy) and K=31 for the high-quality recordings (30 MFCC + log energy). N is the number of speakers used for training, N=5139 for SRE16 DNN and N=7330 for voxceleb DNN.\\
}
\begin{tabular}{c|c|c|c}
Layer &  Frames & Input dim & Output dim \\
       \hline
      \rule[0cm]{-0.1cm}{0.4cm}
       frame-level 1 & 5 & 5*K & 512 \\
      frame-level 2 & 9 & 1536 & 512 \\
    frame-level 3 & 15 & 1536 & 512 \\
    frame-level 4 & 15 & 512 & 512 \\
    frame-level 5 & 15 & 512 & 1500 \\
       pooling  & T & 1500*T & 3000 \\
    segment-level 6 & T & 3000 & 512 \\
   segment-level 7 & T & 512 & 512 \\
    softmax & T & 512 & N 
\end{tabular}
\label{table_DNN}
\end{table}

\paragraph{X-vector extraction}
\label{section_xvect_metho}

In order to extract x-vectors from each subject of our databases we had to extract MFCC in the same way as it was done for the pretrained DNN. We extracted the log energy and 23 MFCC each 10ms for our telephone recordings (like SRE16 model) and 30 MFCC with log energy for our high-quality recordings (like voxceleb model). For the high-quality microphone recordings, we first had to downsampled them to 16kHz (from 96kHz), in order to match the sampling frequency used for the DNN training. Moreover, for this database as for the MFCC-GMM analysis, we carried out spectral subtraction to compensate for mismatched background noises. Voice activity detection and cepstral mean subtraction were also performed on both databases, like SRE16 and voxceleb models and as for our MFCC-GMM analysis. 

X-vectors were then extracted for each subject. They were defined as the 512-dimensional vector extracted after the first segment-level layer of the DNN, just before the non-linear activation function ReLu.

Even if the audio segment tested did not belong to any speaker used to train the DNN, the x-vector extracted can be considered as a representation of this segment and therefore of the speaker.  Back-end analyses can then be carried out to compare and classify the x-vectors according to PD status.

The audio segments used for DNN training had a duration of [2-4s] (after silence removal). This implied compatible segment durations comprised between 25ms to 100s for any speech we wanted to extract x-vectors from. Audio segments with a duration inferior to 25ms would not be not taken into account. Segments longer than 100s would be divided into fragments smaller than 100s. X-vectors corresponding to these fragments would then be averaged.

We assessed the impact of matched segment durations between training and test in section \ref{section_duration}. For all the other experiments we chose to divide our audio files into [1-5s] segments.

\paragraph{Data augmentation}
\label{section_aug_metho}

Recently, enhanced speaker recognition with i-vectors and x-vectors has been noted by augmenting data \cite{snyder_x-vectors:_2018} for DNN and PLDA training. Data augmentation consisted in duplicating the data, adding additive noises and echo to the copies.  Thus, this led to increased quantity and diversity of samples available for the training. In our analyses, data augmentation was performed during DNN training and we assessed its effect on LDA and PLDA training. We used 4 different types of data augmentation: 

- Echo: a reverberation was simulated by taking the convolution of our data with Room Impulse Response (RIR) of different shapes and sizes, available online (\url{http://www.openslr.org/28}).

- Additive noise: different types of noises, extracted from MUSAN database (\url{http://www.openslr.org/17}), were added additively, every second.

- Additive music: musical extracts (from MUSAN database) were added as background noise.

- Babble: three to seven speakers (from MUSAN database) were randomly selected, summed together, then added to our data.

MUSAN and RIR NOISES databases were sampled at 16kHz, we downsampled them to 8kHz for the telephone recordings analysis.

Half of the four augmented copies were finally randomly picked and added to our training database, multiplying by three the size of the latter.

\paragraph{Back-end analyses}

\begin{figure*}[htb]
      \centering
      \includegraphics[scale=0.55]{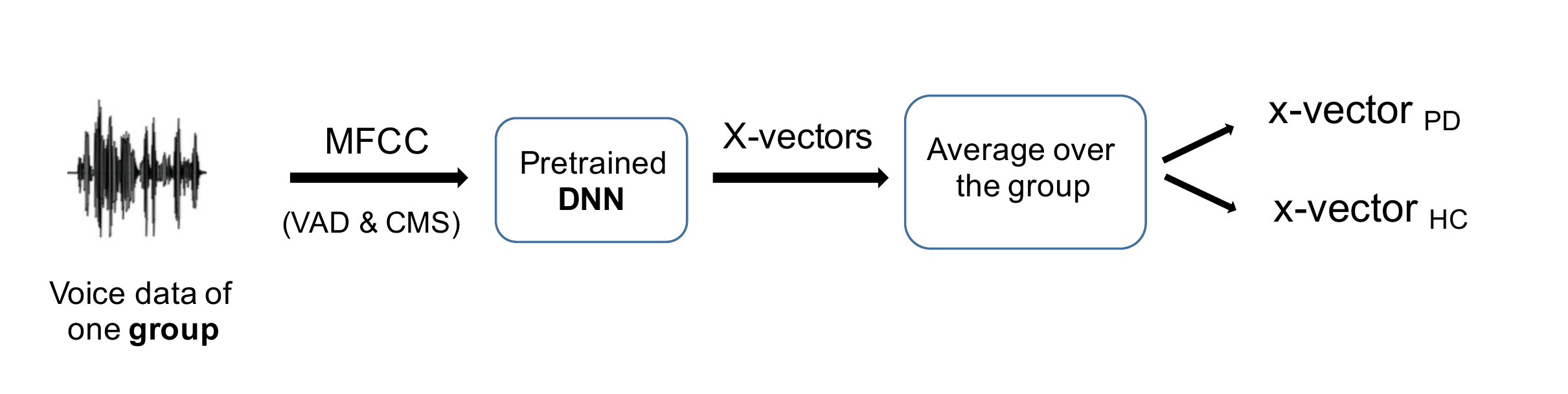}
     \caption{Reference x-vectors: x-vectors are computed for all the training subjects from their MFCC, then averaged within the training groups (male PD, female PD, male control and female control) in order to have one average x-vector per group. VAD: Voice Activity Detection, CMS: Cepstral Mean Subtraction, DNN: Deep Neural Network.} 
     \label{train_xvect}
   \end{figure*}

Once the x-vectors extracted for each subject, x-vectors of PD training group and x-vectors of HC training group were averaged in order to have one average x-vector representing each class, for each gender (see Figure \ref{train_xvect}). 

\begin{figure*}[htb]
      \centering
      \includegraphics[scale=0.55]{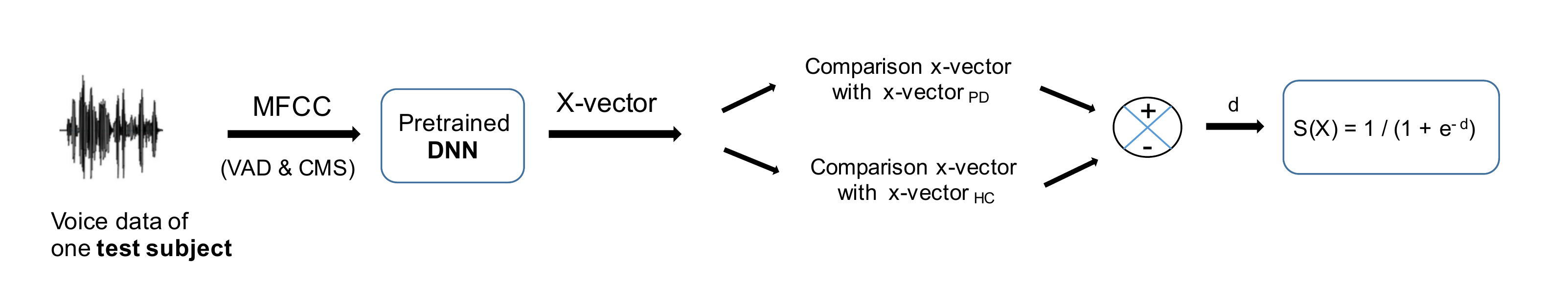}
     \caption{x-vector test phase: x-vectors are computed for each test subject from their MFCC, then compared to the average x-vector$_{PD}$ and  x-vector$_{HC}$. For the comparison we used distance cosine (alone or after LDA projection) and PLDA. The sigmoid of the difference between similarity scores provides the classification score. VAD: Voice Activity Detection, CMS: Cepstral Mean Subtraction, DNN: Deep Neural Network.} 
     \label{schema_xvect}
   \end{figure*}

Classification of test subjects was done by comparing their x-vectors to the average x-vector$_{PD}$ and x-vector$_{HD}$, using a "distance" measure. The difference between these two "distances'' was then calculated and normalized with a sigmoid function, providing a classification score between 0 and 1, per x-vector (see Figure \ref{schema_xvect}). When there were several audio segments for a test subject, i.e. several x-vectors, the average of classification scores of all the x-vectors was performed.
All the participants were split into training and test groups the same way as for MFCC-GMM analysis.

Several methods exist to measure distance between vectors. We compared 3 methods often used with i-vectors or x-vectors: cosine distance, cosine distance preceded by LDA, and PLDA.

\subparagraph{Cosine distance and Linear Discriminant Analysis}

Cosine distance between two vectors is a simple distance measure consisting in calculating the cosine of the angle formed between these two vectors. 

In order to reduce intra-class variability and raise inter-class variability, discriminant analyses may complete the back-end process.  We supplemented the previous cosine distance with a 2-dimensional LDA. LDA training consisted in finding the orthogonal basis onto which the projection of x-vectors (extracted from our training groups) maximized intra-class variability while minimizing inter-class variability. Then cosine distance was computed within this subspace.

\subparagraph{Probabilistic Linear Discriminant Analysis}

Discriminant analysis can also be performed in a probabilistic way. PLDA was introduced in 2007 for face recognition \cite{prince_probabilistic_2007} with i-vectors. We adapted it to PD detection with x-vectors.
We decomposed x-vectors  $\textbf{x}$ into an average component $\bm{\mu}$, computed on all the training subjects, a class-specific part $F.\textbf{h}$, a speaker and session related part $G.\textbf{w}$ and a residual term $\bm{\epsilon}$ assumed to be Gaussian with zero mean and diagonal covariance $\Sigma$ (see Equation \ref{eq_PLDA}).

\begin{equation}
\label{eq_PLDA}
\textbf{x}=\bm{\mu} + F.\textbf{h} + G.\textbf{w} +\bm{\epsilon}
\end{equation}

$F$ matrix columns represent the basis explaining the inter-class variance, with vector $\textbf{h}$ the position of the subject in this subspace. $G$ matrix columns represent the basis explaining the intra-class variance, with vector $\textbf{w}$ the position of the speaker in this subspace.
During the training phase, $\mu$, $F$, $G$ and $\Sigma$ are estimated. During the test phase, x-vectors of test subjects are compared to x-vector$_{PD}$ and x-vector$_{HC}$ by assessing the probability that they share the same identity variable $\textbf{h}$. 

 PLDA was preceded by an LDA in order to reduce the x-vector dimension.

\paragraph{Validation and ensemble method}

For the final classification and the validation we kept the ensemble method used for MFCC-GMM analysis and described in section \ref{section_agg_metho}.

\section{Results and discussion}

In the following section we present the results of x-vector analysis compared to MFCC-GMM for both sexes and for both recording types (high-quality and telephone). We analyzed the effect of audio segment duration, data augmentation, gender, type of classifier (for each speech task), dataset used for DNN training and the choice of an ensemble method. More details about MFCC-GMM analysis (men only) can be found in \cite{jeancolas_comparison_2019}, in particular regarding the comparison of high-quality microphone vs. telephone recordings, as well as speech task effects. 
Performances were measured with the Equal Error Rate (EER), i.e. the error rate corresponding to the threshold for which false positive ratio is equal to false negative ratio (i.e. sensitivity equal to specificity).

\subsection{Impact of segment duration}
\label{section_duration}

 In order to have enough x-vectors for LDA and PLDA training, we segmented our training audio files into [1-5s] segments. For the test phase, we compared two conditions. In the first condition, we considered a large variety of segment durations, from 25ms to 100s (in order to stay in the DNN compatible limits as explained in section \ref{section_xvect_metho}). Those test segments were then neither matched with the duration of segments used for DNN training nor LDA and PLDA training, nor for the constitution of average x-vector$_{PD}$ and x-vector$_{HC}$. In the second condition, we divided all our audio files into [1-5s] segments. Test segment durations were then matched with training segment durations. Results for both duration conditions are presented in Table \ref{table_segment} for the three classification methods (cosine distance alone, with LDA, and PLDA).
 We noticed an improvement of around 3\% for the three classifiers for the [1-5s] test segments. The improvement may be due to matching durations between training segments and test segments, or to the fact that classification was performed on more test segments (because shorter on average). This would compensate for the fact that taken separately, 
long segments have been shown to be better classified than short segments in speaker and language recognition \cite{snyder_deep_2017, snyder_spoken_2018}.
For the next experiments, we kept matched segment durations. 

\begin{table}[!htb]
\centering
\caption{Classification EER (in \%) for male PD vs HC with telephone recordings (sentence repetition task). Comparison of different segment lengths used for x-vectors extraction:  [1-5s] segments for training and either [15ms-100s] (mismatched) or [1-5s] (matched) segments for test.\\
}
\begin{tabular}{l|cc}
Classifier & mismatched  & matched  \\
       \hline
     \rule[0cm]{-0.1cm}{0.4cm}
      x-vec + cos  & 41 & \textbf{39} \\
      x-vec + LDA + cos  & 36 & \textbf{32} \\
      x-vec + PLDA  & 36 & \textbf{33} 
\end{tabular}
\label{table_segment}
\end{table}

\subsection{Impact of data augmentation}

In this section we assessed the impact of augmenting LDA and PLDA training data.
Results obtained with and without data augmentation for LDA and PLDA training are detailed in Table \ref{table_classifier} for the free speech and sentence repetition tasks and in Table \ref{table_DDK} for DDK task. We observed 2-3\% enhancement with data augmentation for the free speech task,  but no consistent improvement for sentence repetition tasks or DDK tasks. This can be explained by the fact that data augmentation added phonetic variability which may have damaged the specificity of the phonetic content of the text-dependent tasks (like sentence repetitions, reading or DDK tasks). Data augmentation seems to be more suited to text-independent tasks (like free speech).

\begin{table*}[tb]
\centering
\caption{Classification EER (in \%) for PD vs HC with high-quality microphone and telephone. Comparison of classifiers: MFCC-GMM (baseline) and x-vectors combined either with cosine distance (alone and with LDA) or with PLDA, and effect of data augmentation. Analyzed tasks are free speech (monologue) and sentence repetitions (combined with readings for high-quality microphone recordings).\\
}
\begin{tabular}{l|cc|cc|cc|cc}
     &   \multicolumn{4}{c|}{High-quality microphone}    & \multicolumn{4}{c}{Telephone}\\
     \cline{2-9}
    \rule[0cm]{0cm}{0.4cm}
     &   \multicolumn{2}{c|}{Males}  & \multicolumn{2}{c|}{Females} & \multicolumn{2}{c|}{Males} & \multicolumn{2}{c}{Females}   \\
   & Repet &  Monol &  Repet &  Monol & Repet &  Monol &  Repet &  Monol \\
       \hline
       \rule[0cm]{-0.1cm}{0.4cm}
      MFCC-GMM    & \textbf{22}  & 26  & 42  & 45  &     35 & 36  & 42  & 40  \\
      x-vec + cos                & 32   & 35  & 51  & 41  &    39  &  33 & 49  & 43 \\
      x-vec + LDA + cos       & \textbf{22}  & 27  & 39  & 32  &     \textbf{32}  & 35  & \textbf{34}  & 34  \\
      x-vec + augLDA + cos  & 24  & \textbf{25}  & \textbf{34}  & \textbf{30}   &    33  & \textbf{33}  & 39  & \textbf{33}  \\
      x-vec + PLDA                & 24  & 28  & 39  & 35  &     33  & 36  & \textbf{34}  & 36 \\
     x-vec +  augPLDA          & 25  & \textbf{25}  & \textbf{33}  & \textbf{30}   &   \textbf{31}  & \textbf{33}  & 37  & \textbf{33} 
\end{tabular}
\label{table_classifier}
\end{table*}

\subsection{Gender effect}

MFCC-GMM and x-vector classifiers were trained separately for each gender, in order to study gender effect on early PD detection. For all classifiers we noticed an important gender effect with better performances for male PD detection (see Table \ref{table_classifier}). Several reasons may explain these gender differences. 

First of all, previous studies have reported wider female MFCC distribution with more variability, making MFCC based classifications more difficult in women \cite{fraile_automatic_2009}. The authors of \cite{tsanas_nonlinear_2011} also noticed that MFCC features were more suited to monitor PD evolution in men than women. This may explain the poor classification performances with MFCC-GMM method in women. 

Interestingly, x-vectors when combined with discriminant analysis (LDA or PLDA) clearly improved female classification performances. This was certainly due to the fact that these discriminant analyses reduced intra-class variance, and thus tackled the MFCC variability issue in women.
LDA and PLDA reduced the classification performance gap between genders but did not suppress it entirely. The remaining differences may be explained by other factors. 

First, less pronounced brain atrophy \cite{tremblay_gender_2019} and less network disruptions \cite {haaxma_gender_2007} have been observed in the first stages of PD in women. In addition, the onset of symptoms is delayed on average by two years in women compared to men \cite {haaxma_gender_2007}. A possible protective role of estrogen on PD has often been suggested to explain gender differences in early PD manifestations. Besides we can notice in our age-matched database a lower UPDRS III motor score in PD women compared PD men (see Table \ref{table_database_pro} and \ref{table_database_tel}). A second factor possibly leading to gender differences in PD detection through voice, is that speech neural circuits are different in men and women \cite{de_lima_xavier_sexual_2019, jung_sex_2019}. These circuits may therefore be differently affected in PD, and lead to different types or degrees of vocal impairments.

\subsection{Comparison of classifiers and influence of speech task type}

In this section we compared the different classification methodologies using x-vectors among themselves and with MFCC-GMM classification. 
First, we observed that cosine distance combined with LDA performs as well as PLDA, and globally better than cosine distance alone, whatever the recording condition (telephone or high-quality microphone) or speech task (Table \ref{table_classifier} and \ref{table_DDK}). This improvement due to discriminant analysis was encountered in both gender but was sharper in women (as explained previously). 

We already showed that data augmentation for LDA and PLDA training improved classification for the free speech task but not for the text-dependent tasks. Therefore, for the comparison between MFCC-GMM  and x-vectors, we used for the latter, cosine distance combined with augmented LDA for free speech task, and not augmented LDA for sentence repetitions and DDK tasks. 

\begin{figure*}[htb]
      \centering
      \includegraphics[scale=1.8]{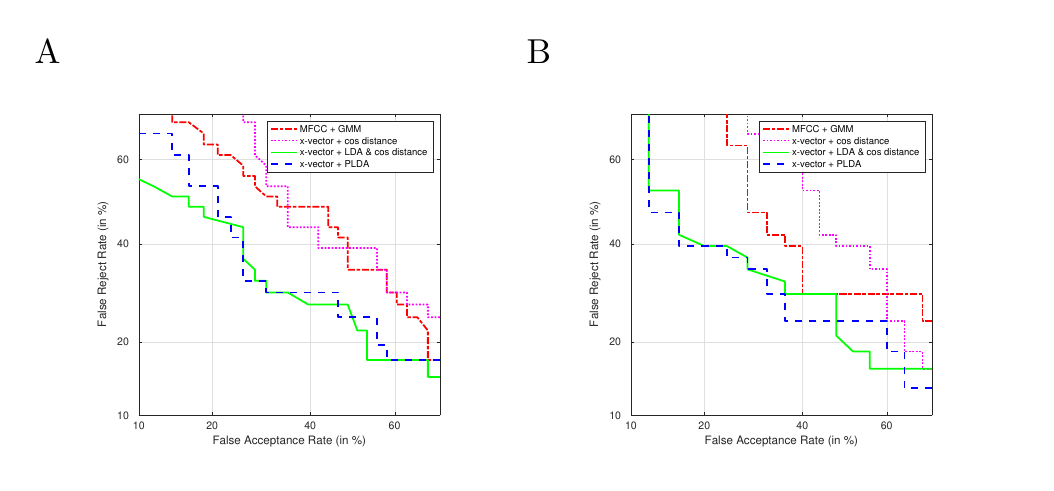}
     \caption{DET curves of female classification PD vs HC, using free speech task, recorded with the high-quality microphone (A) and with their telephone (B). Comparison of classifiers performances: MFCC-GMM (baseline) and x-vectors combined either with cosine distance (alone and with LDA) or with PLDA. LDA and PLDA are performed with data augmentation. }
     \label{DET_femmes}
   \end{figure*}

For all recording conditions and both genders, we observed improved classification performances with x-vectors (compared to MFCC-GMM) for the free speech task (see Table \ref{table_classifier}). This is consistent with the fact that x-vectors were originally developed for text-independent speaker recognition. This improvement with x-vectors was even more pronounced in women (7\% increase with telephone and 15\% with high-quality microphone). Detection Error Tradeoff (DET) curves in Figure \ref{DET_femmes} illustrate this classifier comparison in women.

An overall improvement with x-vectors also appeared for sentence repetitions and readings but in a less marked way.

Finally, very specific tasks, such as DDK (tested on men), performed better with MFCC-GMM than with x-vectors (see Table \ref{table_DDK}). This could be due to the DNN we used to extract x-vectors, that was trained on conversations, containing wider variety of phonemes than in DDK tasks (composed of vowels and stop consonants only). Thus, DDK specificity was not exploited by the calibration of the DNN, resulting in a loss of discriminating power with x-vectors.

\subsection{Comparison with DNN trained with our own data}
\label{section_ourDNN}

In order to make the DNN more suitable for the particular type of DDK tasks, we carried out an additional experiment, training this time the DNN with DDK tasks of our own data.
The subjects used for DNN training were the same as those used for the constitution of the average x-vector$_{PD}$ and x-vector$_{HC}$ and the LDA and PLDA training. Remaining subjects were used for the test. The results obtained are presented in Table \ref{table_DDK}. 
We noticed a clear performance degradation when data augmentation was applied on LDA and PLDA training. This is consistent with the fact that data augmentation, while adding noises, impairs the specificity of the DDK phonetic content.

Results obtained with cosine distance + LDA and PLDA, without data augmentation, were similar to those obtained with the previous pretrained DNN. Our DNN training was certainly more specific but perhaps suffered from insufficient data quantity, which could explain why it did not lead to better performance.

\begin{table}[!htb]
\centering
\caption{Classification EER (in \%) for male PD vs HC with telephone recordings (DDK task). Comparison of two databases used for DNN training, SRE16 database and our telephone database. \\
}
\begin{tabular}{l|cc}
Classifier     &   SRE16   DNN    & our DNN\\
       \hline
       \rule[0cm]{-0.1cm}{0.4cm}
      MFCC-GMM  & \textbf{25}  & \textbf{25}  \\
       x-vec + cos  & 35  & 47  \\
     x-vec + LDA + cos  & 29  & 29   \\
     x-vec + augLDA + cos  & 30  & 39   \\
    x-vec +  PLDA  & 30  & 30  \\
     x-vec + augPLDA  & 30  & 38  
\end{tabular}
\label{table_DDK}
\end{table}

\subsection{Aggregated model vs simple model}
\label{section_agg_result}

In order to test the advantage of the ensemble method we used, we compared its performances with the results obtained with the related simple model.
To estimate the performance of the simple model, we performed a classic random subsampling cross validation. We averaged the DET curves from each run and calculated the EER corresponding to the average DET curve. The performances obtained are detailed in Table \ref{table_agg} and compared to an extract of Table \ref{table_classifier}. With both MFCC-GMM and x-vector classifiers we observed a 2-3\% improvement for the aggregated model, compared to the simple model. This demonstrates the interest of using ensemble methods for PD detection using voice.

\begin{table}[!htb]
\centering
\caption{Classification EER (in \%) for male PD vs HC with telephone recordings. Most appropriate tasks (diadochokinesia, sentence repetition or monologue) are used for each classifier. Comparison of the aggregated model (ensemble method) with the simple model. \\
}
\begin{tabular}{l|ccc}
Classifier     &   Task    & Aggregated & Simple\\
\hline
\rule[0cm]{-0.1cm}{0.4cm}
      MFCC-GMM  & DDK & \textbf{25}  & 28  \\
      x-vec + LDA + cos  & Repet & \textbf{32}  & 35   \\
     x-vec + augLDA + cos  & Monol & \textbf{33}  & 35   \\
     x-vec + PLDA  & Repet & \textbf{33}  & 35  \\
    x-vec +  augPLDA  & Monol & \textbf{33}  & 35  
\end{tabular}
\label{table_agg}
\end{table}

\section{Conclusion}

According to the literature, the latest speaker recognition system, called x-vectors, provides more robust speaker representations and enhanced recognition, when large amount of training data is used. Our goal was to assess if this technique could be adapted to early PD detection (from recordings done with a high-quality microphone and by telephone) and improve the detection performances. We compared x-vector classification method to a more classic system based on MFCC and GMM.

At first, we recorded 221 French speakers (including PD subjects recently diagnosed and healthy controls) with a high-quality microphone and with their telephone. Our voice analyses were based on MFCC features. The baseline consisted in modeling PD and HC distribution with GMM. For x-vector technique, MFCC were used as inputs of a  feed-forward DNN from which embeddings (called x-vectors) were extracted then classified. Since DNN training usually requires a lot of data, we used a DNN trained on large speaker recognition databases. All the analyses were done in a separate way for men and women, in order to avoid additional variability due to gender, and to study a possible gender effect on early PD detection. We varied several experimental and methodological aspects in order to analyze their effect on the classification performances. 

\noindent \textbf{Influence of segment duration:}
We observed that using short audio segments that were matched between training and test provided better results. \\
\textbf{Comparison of back-end analyses:}
We compared different back-end analyses used with x-vectors. We noticed that the addition of LDA clearly improved the cosine distance classification and performed as well as a PLDA classifier. This improvement due to discriminant analyses was even more pronounced in women, whose voices are known to contain more variability.\\
\textbf{Influence of data augmentation:}
We found that augmenting data for the training of LDA and PLDA led to improved classification for the free speech task but not for text-dependent tasks (like sentence repetitions and DDK). This is consistent with the fact that adding noised data copies increases quantity but impairs the specificity of phonetic contents. \\
\textbf{Comparison MFCC-GMM vs. x-vectors for different speech task types:}
The comparison with MFCC-GMM classification showed that x-vectors performed better for the free-speech task, which is consistent with the fact that x-vectors were originally developed for text-independent speaker recognition. Very specific tasks, like DDK, resulted in better performances with GMM. Lower results with x-vectors for this task may be due to the varied phonetic content used to pretrain the DNN, whereas the GMM were trained with our DDK data, thus preserving the speech task specificity.\\
\textbf{Gender effect:}
We noticed lower performances in PD detection in women compared to men, with MFCC-GMM. This is consistent with a higher MFCC variability in women. X-vectors combined with LDA or PLDA handled this variability and led to 7 to 15\% classification improvement. Differences between speech neural circuits in men and women and a disease less pronounced in women at the first stages may explain the remaining classification performance differences.\\
\textbf{Influence of dataset used for DNN training:}
 In order to make the DNN more specific to DDK tasks, we carried out an additional analysis by training it this time with our database (from DDK tasks). The performances obtained were not improved compared to the pretrained DNN, showing the importance of data quantity on DNN training. \\
\textbf{Influence of ensemble method:}
 Finally, we observed a 2-3\% classification improvement when the ensemble method was used, for both MFCC-GMM and x-vectors classifiers.
 
To conclude, x-vectors, combined with discriminant analyses, seems to be more relevant than MFCC-GMM classification for text-independent tasks and particularly suited to women PD detection.

In future work, we will study features related to other PD vocal disruptions, like phonation, prosody and rhythmic abilities and combine them with this analysis (more related to articulation disorder) in order to gather all the information we can have on early PD voice and enhance the detection.

\section*{Conflict of Interest Statement}

The authors declare that the research was conducted in the absence of any commercial or financial relationships that could be construed as a potential conflict of interest.

\section*{Author Contributions}

LJ: experimental design, data collection, data analysis and interpretation, and manuscript draft. 
DP: experimental design, validation of the analysis and its interpretation, manuscript revision.
GM: participants' diagnosis and clinical scores.
BB:  validation of the analysis and its interpretation.
JC, MV, SL: design and development of ICEBERG study, data collection and manuscript revision.
HB: validation of the analysis and its interpretation, and manuscript revision.

\section*{Funding}
L. Jeancolas was supported by a grant of Institut Mines-Télécom, Fondation Télécom and Institut Carnot Télécom \&
Société Numérique through ``Futur \& Ruptures" program. The
ICEBERG study was partly funded by the program ``Investissements
d’Avenir” ANR-10-IAIHU-06 (Paris Institute of Neurosciences – IHU), ANR-11-INBS-0006, Fondation EDF, Fondation Planiol, Société Française de Médecine Esthétique (Mr. Legrand) and Energipole (Mr. Mallard).

\section*{Acknowledgments}
 The authors would like to thank Samovar laboratory (especially Mohamed Amine Hmani and Aymen Mtibaa), CIC Neurosciences (especially Alizé Chalançon, Christelle Laganot and Sandrine Bataille), sleep disorder unit and CENIR teams. The authors are also grateful to Obaï Bin Ka'b Ali and Fatemeh Razavipour for the manuscript revision. Finally, the authors would like to express their sincere acknowledgments to all the subjects who have participated in this study.

\section*{Data Availability Statement}
The datasets presented in this article are not readily available in order to comply with the ethical consents provided by the participants. Requests to access the datasets should be directed to Marie Vidailhet (marie.vidailhet@aphp.fr).

\bibliography{bibliothese}

\begin{thebibliography}{65}
\providecommand{\natexlab}[1]{#1}
\providecommand{\url}[1]{\texttt{#1}}
\expandafter\ifx\csname urlstyle\endcsname\relax
  \providecommand{\doi}[1]{doi: #1}\else
  \providecommand{\doi}{doi: \begingroup \urlstyle{rm}\Url}\fi

\bibitem[Arora et~al.(2019)Arora, Baghai-Ravary, and
  Tsanas]{arora_developing_2019}
S.~Arora, L.~Baghai-Ravary, and A.~Tsanas.
\newblock Developing a large scale population screening tool for the assessment
  of {Parkinson}'s disease using telephone-quality voice.
\newblock \emph{The Journal of the Acoustical Society of America}, 145\penalty0
  (5):\penalty0 2871--2884, May 2019.
\newblock ISSN 0001-4966.
\newblock \doi{10.1121/1.5100272}.
\newblock URL \url{http://asa.scitation.org/doi/10.1121/1.5100272}.

\bibitem[Benba et~al.(2014)Benba, Jilbab, and Hammouch]{benba_voice_2014}
A.~Benba, A.~Jilbab, and A.~Hammouch.
\newblock Voice analysis for detecting persons with {Parkinson}’s disease
  using {MFCC} and {VQ}.
\newblock In \emph{The 2014 international conference on circuits, systems and
  signal processing}, pages 23--25, 2014.
\newblock URL
  \url{http://www.inase.org/library/2014/russia/bypaper/CCCS/CCCS-15.pdf}.

\bibitem[Benba et~al.(2016)Benba, Jilbab, and
  Hammouch]{benba_discriminating_2016}
A.~Benba, A.~Jilbab, and A.~Hammouch.
\newblock Discriminating {Between} {Patients} {With} {Parkinson}'s and
  {Neurological} {Diseases} {Using} {Cepstral} {Analysis}.
\newblock \emph{IEEE Transactions on Neural Systems and Rehabilitation
  Engineering}, 24\penalty0 (10):\penalty0 1100--1108, Oct. 2016.
\newblock ISSN 1534-4320.
\newblock \doi{10.1109/TNSRE.2016.2533582}.

\bibitem[Bimbot et~al.(2004)Bimbot, Bonastre, Fredouille, Gravier,
  Magrin-Chagnolleau, Meignier, Merlin, Ortega-García, Petrovska-Delacrétaz,
  and Reynolds]{bimbot_tutorial_2004}
F.~Bimbot, J.-F. Bonastre, C.~Fredouille, G.~Gravier, I.~Magrin-Chagnolleau,
  S.~Meignier, T.~Merlin, J.~Ortega-García, D.~Petrovska-Delacrétaz, and
  D.~A. Reynolds.
\newblock A {Tutorial} on {Text}-{Independent} {Speaker} {Verification}.
\newblock \emph{EURASIP Journal on Advances in Signal Processing},
  2004\penalty0 (4):\penalty0 101962, Dec. 2004.
\newblock ISSN 1687-6180.
\newblock \doi{10.1155/S1110865704310024}.
\newblock URL
  \url{https://asp-eurasipjournals.springeropen.com/articles/10.1155/S1110865704310024}.

\bibitem[Bocklet et~al.(2013)Bocklet, Steidl, Nöth, and
  Skodda]{bocklet_automatic_2013}
T.~Bocklet, S.~Steidl, E.~Nöth, and S.~Skodda.
\newblock Automatic evaluation of parkinson's speech-acoustic, prosodic and
  voice related cues.
\newblock In \emph{Interspeech}, pages 1149--1153, 2013.
\newblock URL
  \url{http://ai2-s2-pdfs.s3.amazonaws.com/ecfe/10780a4e45031024860068d8cc98b78abb44.pdf}.

\bibitem[Boersma and Weenink(2001)]{boersma_praat_2001}
P.~Boersma and D.~Weenink.
\newblock {PRAAT}, a system for doing phonetics by computer.
\newblock \emph{Glot international}, 5:\penalty0 341--345, Jan. 2001.

\bibitem[Boll(1979)]{boll_suppression_1979}
S.~Boll.
\newblock Suppression of acoustic noise in speech using spectral subtraction.
\newblock \emph{IEEE Transactions on Acoustics, Speech, and Signal Processing},
  27\penalty0 (2):\penalty0 113--120, Apr. 1979.
\newblock ISSN 0096-3518.
\newblock \doi{10.1109/TASSP.1979.1163209}.

\bibitem[Breiman(1996)]{breiman_bagging_1996}
L.~Breiman.
\newblock Bagging predictors.
\newblock \emph{Machine Learning}, 24\penalty0 (2):\penalty0 123--140, Aug.
  1996.
\newblock ISSN 0885-6125, 1573-0565.
\newblock \doi{10.1007/BF00058655}.
\newblock URL \url{http://link.springer.com/10.1007/BF00058655}.

\bibitem[Bühlmann and Yu(2002)]{buhlmann_analyzing_2002}
P.~Bühlmann and B.~Yu.
\newblock Analyzing {Bagging}.
\newblock \emph{The Annals of Statistics}, 30\penalty0 (4):\penalty0 927--961,
  2002.

\bibitem[De~Lau and Breteler(2006)]{de_lau_epidemiology_2006}
L.~M. De~Lau and M.~M. Breteler.
\newblock Epidemiology of {Parkinson}'s disease.
\newblock \emph{The Lancet Neurology}, 5\penalty0 (6):\penalty0 525--535, 2006.
\newblock URL
  \url{http://www.sciencedirect.com/science/article/pii/S1474442206704719}.

\bibitem[de~Lima~Xavier et~al.(2019)de~Lima~Xavier, Hanekamp, and
  Simonyan]{de_lima_xavier_sexual_2019}
L.~de~Lima~Xavier, S.~Hanekamp, and K.~Simonyan.
\newblock Sexual {Dimorphism} {Within} {Brain} {Regions} {Controlling} {Speech}
  {Production}.
\newblock \emph{Frontiers in Neuroscience}, 13, 2019.
\newblock ISSN 1662-453X.
\newblock \doi{10.3389/fnins.2019.00795}.
\newblock URL
  \url{https://www.frontiersin.org/articles/10.3389/fnins.2019.00795/full}.

\bibitem[Dibazar et~al.(2002)Dibazar, Narayanan, and
  Berger]{dibazar_feature_2002}
A.~A. Dibazar, S.~Narayanan, and T.~W. Berger.
\newblock Feature analysis for automatic detection of pathological speech.
\newblock In \emph{Proceedings of the {Second} {Joint} 24th {Annual}
  {Conference} and the {Annual} {Fall} {Meeting} of the {Biomedical}
  {Engineering} {Society}] [{Engineering} in {Medicine} and {Biology}},
  volume~1, pages 182--183 vol.1, 2002.
\newblock \doi{10.1109/IEMBS.2002.1134447}.

\bibitem[Ene(2008)]{ene_neural_2008}
M.~Ene.
\newblock Neural network-based approach to discriminate healthy people from
  those with {Parkinson}’s disease.
\newblock \emph{Annals of the University of Craiova, Mathematics and Computer
  Science}, 35:\penalty0 112--116, 2008.

\bibitem[Fearnley and Lees(1991)]{fearnley_ageing_1991}
J.~M. Fearnley and A.~J. Lees.
\newblock Ageing and {Parkinson}'s disease: substantia nigra regional
  selectivity.
\newblock \emph{Brain: A Journal of Neurology}, 114 ( Pt 5):\penalty0
  2283--2301, Oct. 1991.
\newblock ISSN 0006-8950.

\bibitem[Fraile et~al.(2009)Fraile, Sáenz-Lechón, Godino-Llorente, Osma-Ruiz,
  and Fredouille]{fraile_automatic_2009}
R.~Fraile, N.~Sáenz-Lechón, J.~Godino-Llorente, V.~Osma-Ruiz, and
  C.~Fredouille.
\newblock Automatic {Detection} of {Laryngeal} {Pathologies} in {Records} of
  {Sustained} {Vowels} by {Means} of {Mel}-{Frequency} {Cepstral} {Coefficient}
  {Parameters} and {Differentiation} of {Patients} by {Sex}.
\newblock \emph{Folia Phoniatrica et Logopaedica}, 61\penalty0 (3):\penalty0
  146--152, July 2009.
\newblock ISSN 1021-7762, 1421-9972.
\newblock \doi{10.1159/000219950}.
\newblock URL \url{http://www.karger.com/?doi=10.1159/000219950}.

\bibitem[Friedman et~al.(2001)Friedman, Hastie, and
  Tibshirani]{friedman_elements_2001}
J.~Friedman, T.~Hastie, and R.~Tibshirani.
\newblock \emph{The elements of statistical learning}, volume~1.
\newblock Springer series in statistics Springer, Berlin, 2001.
\newblock URL \url{http://statweb.stanford.edu/~tibs/book/preface.ps}.

\bibitem[Garcia et~al.(2017)Garcia, Vásquez-Correa, Orozco-Arroyave, Dehak,
  and Nöth]{garcia_language_2017}
N.~Garcia, J.~C. Vásquez-Correa, J.~R. Orozco-Arroyave, N.~Dehak, and
  E.~Nöth.
\newblock Language {Independent} {Assessment} of {Motor} {Impairments} of
  {Patients} with {Parkinson}’s {Disease} {Using} i-{Vectors}.
\newblock In \emph{Text, {Speech}, and {Dialogue}}, volume 10415, pages
  147--155, Cham, 2017. Springer International Publishing.
\newblock ISBN 978-3-319-64205-5 978-3-319-64206-2.
\newblock \doi{10.1007/978-3-319-64206-2_17}.
\newblock URL \url{http://link.springer.com/10.1007/978-3-319-64206-2_17}.

\bibitem[Garcia-Ospina et~al.(2018)Garcia-Ospina, Arias-Vergara,
  Vásquez-Correa, Orozco-Arroyave, Cernak, and
  Nöth]{garcia-ospina_phonological_2018}
N.~Garcia-Ospina, T.~Arias-Vergara, J.~C. Vásquez-Correa, J.~R.
  Orozco-Arroyave, M.~Cernak, and E.~Nöth.
\newblock Phonological i-{Vectors} to {Detect} {Parkinson}’s {Disease}.
\newblock In P.~Sojka, A.~Horák, I.~Kopeček, and K.~Pala, editors,
  \emph{Text, {Speech}, and {Dialogue}}, Lecture {Notes} in {Computer}
  {Science}, pages 462--470. Springer International Publishing, 2018.
\newblock ISBN 978-3-030-00794-2.

\bibitem[Gil and Johnson(2009)]{gil_diagnosing_2009}
D.~Gil and M.~Johnson.
\newblock Diagnosing {Parkinson} by using {Artificial} {Neural} {Networks} and
  {Support} {Vector} {Machines}.
\newblock \emph{Global Journal of Computer Science and Technology}, 9, 2009.

\bibitem[Godino-Llorente and
  Gómez-Vilda(2004)]{godino-llorente_automatic_2004}
J.~Godino-Llorente and P.~Gómez-Vilda.
\newblock Automatic {Detection} of {Voice} {Impairments} by {Means} of
  {Short}-{Term} {Cepstral} {Parameters} and {Neural} {Network} {Based}
  {Detectors}.
\newblock \emph{IEEE Transactions on Biomedical Engineering}, 51\penalty0
  (2):\penalty0 380--384, Feb. 2004.
\newblock ISSN 0018-9294.
\newblock \doi{10.1109/TBME.2003.820386}.
\newblock URL \url{http://ieeexplore.ieee.org/document/1262116/}.

\bibitem[Goetz et~al.(2007)Goetz, Fahn, Martinez‐Martin, Poewe, Sampaio,
  Stebbins, Stern, Tilley, Dodel, Dubois, Holloway, Jankovic, Kulisevsky, Lang,
  Lees, Leurgans, LeWitt, Nyenhuis, Olanow, Rascol, Schrag, Teresi, Hilten, and
  LaPelle]{goetz_movement_2007}
C.~G. Goetz, S.~Fahn, P.~Martinez‐Martin, W.~Poewe, C.~Sampaio, G.~T.
  Stebbins, M.~B. Stern, B.~C. Tilley, R.~Dodel, B.~Dubois, R.~Holloway,
  J.~Jankovic, J.~Kulisevsky, A.~E. Lang, A.~Lees, S.~Leurgans, P.~A. LeWitt,
  D.~Nyenhuis, C.~W. Olanow, O.~Rascol, A.~Schrag, J.~A. Teresi, J.~J.~V.
  Hilten, and N.~LaPelle.
\newblock Movement {Disorder} {Society}-sponsored revision of the {Unified}
  {Parkinson}'s {Disease} {Rating} {Scale} ({MDS}-{UPDRS}): {Process}, format,
  and clinimetric testing plan.
\newblock \emph{Movement Disorders}, 22\penalty0 (1):\penalty0 41--47, 2007.
\newblock ISSN 1531-8257.
\newblock \doi{10.1002/mds.21198}.
\newblock URL \url{https://onlinelibrary.wiley.com/doi/abs/10.1002/mds.21198}.

\bibitem[Haas et~al.(2012)Haas, Stewart, and Zhang]{haas_premotor_2012}
B.~R. Haas, T.~H. Stewart, and J.~Zhang.
\newblock Premotor biomarkers for {Parkinson}’s disease-a promising direction
  of research.
\newblock \emph{Transl Neurodegener}, 1\penalty0 (1):\penalty0 11, 2012.
\newblock URL
  \url{http://www.biomedcentral.com/content/pdf/2047-9158-1-11.pdf}.

\bibitem[Haaxma et~al.(2007)Haaxma, Bloem, Borm, Oyen, Leenders, Eshuis, Booij,
  Dluzen, and Horstink]{haaxma_gender_2007}
C.~A. Haaxma, B.~R. Bloem, G.~F. Borm, W.~J.~G. Oyen, K.~L. Leenders,
  S.~Eshuis, J.~Booij, D.~E. Dluzen, and M.~W. I.~M. Horstink.
\newblock Gender differences in {Parkinson}'s disease.
\newblock \emph{Journal of Neurology, Neurosurgery \&amp; Psychiatry},
  78\penalty0 (8):\penalty0 819--824, Aug. 2007.
\newblock ISSN 0022-3050.
\newblock \doi{10.1136/jnnp.2006.103788}.
\newblock URL \url{http://jnnp.bmj.com/cgi/doi/10.1136/jnnp.2006.103788}.

\bibitem[Hemmerling et~al.(2016)Hemmerling, Orozco-Arroyave, Skalski, Gajda,
  and Nöth]{hemmerling_automatic_2016}
D.~Hemmerling, J.~R. Orozco-Arroyave, A.~Skalski, J.~Gajda, and E.~Nöth.
\newblock Automatic {Detection} of {Parkinson}’s {Disease} {Based} on
  {Modulated} {Vowels}.
\newblock In \emph{{INTERSPEECH}}, pages 1190--1194, Sept. 2016.
\newblock \doi{10.21437/Interspeech.2016-1062}.

\bibitem[Hoehn and Yahr(1967)]{hoehn_parkinsonism:_1967}
M.~Hoehn and M.~D. Yahr.
\newblock Parkinsonism: onset, progression and mortality.
\newblock \emph{Neurology}, 17\penalty0 (5):\penalty0 427--442, 1967.

\bibitem[Jafari(2013)]{jafari_classification_2013}
A.~Jafari.
\newblock Classification of {Parkinson}'s {Disease} {Patients} using
  {Nonlinear} {Phonetic} {Features} and {Mel}-{Frequency} {Cepstral}
  {Analysis}.
\newblock \emph{Biomedical Engineering: Applications, Basis and
  Communications}, 25\penalty0 (04):\penalty0 1350001, Aug. 2013.
\newblock ISSN 1016-2372, 1793-7132.
\newblock \doi{10.4015/S1016237213500014}.
\newblock URL
  \url{http://www.worldscientific.com/doi/abs/10.4015/S1016237213500014}.

\bibitem[Jeancolas(2019)]{jeancolas_detection_2019}
L.~Jeancolas.
\newblock \emph{Détection précoce de la maladie de {Parkinson} par l'analyse
  de la voix et corrélations avec la neuroimagerie}.
\newblock phdthesis, Université Paris-Saclay, Dec. 2019.
\newblock URL \url{https://tel.archives-ouvertes.fr/tel-02470759}.

\bibitem[Jeancolas et~al.(2016)Jeancolas, Petrovska-Delacrétaz, Lehéricy,
  Benali, and Benkelfat]{jeancolas_analyse_2016}
L.~Jeancolas, D.~Petrovska-Delacrétaz, S.~Lehéricy, H.~Benali, and B.-E.
  Benkelfat.
\newblock L'analyse de la voix comme outil de diagnostic précoce de la maladie
  de {Parkinson} : état de l'art.
\newblock In \emph{{CORESA} 2016 : 18e {Edition} {COmpressions} et
  {REprésentation} des {Signaux} {Audiovisuels}}, pages 113--121, Nancy, May
  2016. CNRS.
\newblock URL
  \url{https://projet.liris.cnrs.fr/coresa/articles/2016/Coresa2016_proceedings.pdf}.

\bibitem[Jeancolas et~al.(2019)Jeancolas, Mangone, Corvol, Vidailhet,
  Lehéricy, Benkelfat, Benali, and
  Petrovska-Delacrétaz]{jeancolas_comparison_2019}
L.~Jeancolas, G.~Mangone, J.-C. Corvol, M.~Vidailhet, S.~Lehéricy, B.-E.
  Benkelfat, H.~Benali, and D.~Petrovska-Delacrétaz.
\newblock Comparison of {Telephone} {Recordings} and {Professional}
  {Microphone} {Recordings} for {Early} {Detection} of {Parkinson}’s
  {Disease}, {Using} {Mel}-{Frequency} {Cepstral} {Coefficients} with
  {Gaussian} {Mixture} {Models}.
\newblock In \emph{Interspeech 2019}, pages 3033--3037. ISCA, Sept. 2019.
\newblock \doi{10.21437/Interspeech.2019-2825}.
\newblock URL
  \url{http://www.isca-speech.org/archive/Interspeech_2019/abstracts/2825.html}.

\bibitem[Jung et~al.(2019)Jung, Mody, Fujioka, Kimura, Okazawa, and
  Kosaka]{jung_sex_2019}
M.~Jung, M.~Mody, T.~Fujioka, Y.~Kimura, H.~Okazawa, and H.~Kosaka.
\newblock Sex {Differences} in {White} {Matter} {Pathways} {Related} to
  {Language} {Ability}.
\newblock \emph{Frontiers in Neuroscience}, 13, 2019.
\newblock ISSN 1662-453X.
\newblock \doi{10.3389/fnins.2019.00898}.
\newblock URL
  \url{https://www.frontiersin.org/articles/10.3389/fnins.2019.00898/full}.

\bibitem[Kapoor and Sharma(2011)]{kapoor_parkinsons_2011}
T.~Kapoor and R.~K. Sharma.
\newblock Parkinson’s disease diagnosis using {Mel}-frequency cepstral
  coefficients and vector quantization.
\newblock \emph{International Journal of Computer Applications}, 14\penalty0
  (3):\penalty0 43--46, 2011.

\bibitem[Khojasteh et~al.(2018)Khojasteh, Viswanathan, Aliahmad, Ragnav, Zham,
  and Kumar]{khojasteh_parkinsons_2018}
P.~Khojasteh, R.~Viswanathan, B.~Aliahmad, S.~Ragnav, P.~Zham, and D.~K. Kumar.
\newblock Parkinson's {Disease} {Diagnosis} {Based} on {Multivariate} {Deep}
  {Features} of {Speech} {Signal}.
\newblock In \emph{2018 {IEEE} {Life} {Sciences} {Conference} ({LSC})}, pages
  187--190, Oct. 2018.
\newblock \doi{10.1109/LSC.2018.8572136}.

\bibitem[Little et~al.(2009)Little, McSharry, Hunter, Spielman, and
  Ramig]{little_suitability_2009}
M.~Little, P.~McSharry, E.~Hunter, J.~Spielman, and L.~Ramig.
\newblock Suitability of {Dysphonia} {Measurements} for {Telemonitoring} of
  {Parkinson}\&amp;\#x0027;s {Disease}.
\newblock \emph{IEEE Transactions on Biomedical Engineering}, 56\penalty0
  (4):\penalty0 1015--1022, Apr. 2009.
\newblock ISSN 0018-9294, 1558-2531.
\newblock \doi{10.1109/TBME.2008.2005954}.
\newblock URL
  \url{http://ieeexplore.ieee.org/lpdocs/epic03/wrapper.htm?arnumber=4636708}.

\bibitem[López et~al.(2019)López, Orozco-Arroyave, and
  Gosztolya]{lopez_assessing_2019}
J.~V.~E. López, J.~R. Orozco-Arroyave, and G.~Gosztolya.
\newblock Assessing {Parkinson}’s {Disease} from {Speech} {Using} {Fisher}
  {Vectors}.
\newblock In \emph{Interspeech 2019}, pages 3063--3067. ISCA, Sept. 2019.
\newblock \doi{10.21437/Interspeech.2019-2217}.
\newblock URL
  \url{http://www.isca-speech.org/archive/Interspeech_2019/abstracts/2217.html}.

\bibitem[Maillard et~al.(2017)Maillard, Arlot, and
  Lerasle]{maillard_cross-validation_2017}
G.~Maillard, S.~Arlot, and M.~Lerasle.
\newblock Cross-validation improved by aggregation: {Agghoo}.
\newblock \emph{hal}, page~21, 2017.

\bibitem[Malyska et~al.(2005)Malyska, Quatieri, and
  Sturim]{malyska_automatic_2005}
N.~Malyska, T.~F. Quatieri, and D.~Sturim.
\newblock Automatic dysphonia recognition using biologically-inspired
  amplitude-modulation features.
\newblock In \emph{Acoustics, {Speech}, and {Signal} {Processing}, 2005.
  {Proceedings}.({ICASSP}'05). {IEEE} {International} {Conference} on},
  volume~1, pages I--873. IEEE, 2005.
\newblock URL \url{http://ieeexplore.ieee.org/abstract/document/1415253/}.

\bibitem[Moro-Velazquez et~al.(2020)Moro-Velazquez, Villalba, and
  Dehak]{moro-velazquez_using_2020}
L.~Moro-Velazquez, J.~Villalba, and N.~Dehak.
\newblock Using {X}-{Vectors} to {Automatically} {Detect} {Parkinson}’s
  {Disease} from {Speech}.
\newblock In \emph{{ICASSP} 2020 - 2020 {IEEE} {International} {Conference} on
  {Acoustics}, {Speech} and {Signal} {Processing} ({ICASSP})}, pages
  1155--1159, May 2020.
\newblock \doi{10.1109/ICASSP40776.2020.9053770}.
\newblock ISSN: 2379-190X.

\bibitem[Moro-Velázquez et~al.(2018)Moro-Velázquez, Gómez-García,
  Godino-Llorente, Villalba, Orozco-Arroyave, and
  Dehak]{moro-velazquez_analysis_2018}
L.~Moro-Velázquez, J.~A. Gómez-García, J.~I. Godino-Llorente, J.~Villalba,
  J.~R. Orozco-Arroyave, and N.~Dehak.
\newblock Analysis of speaker recognition methodologies and the influence of
  kinetic changes to automatically detect {Parkinson}'s {Disease}.
\newblock \emph{Applied Soft Computing}, 62:\penalty0 649--666, Jan. 2018.
\newblock ISSN 15684946.
\newblock \doi{10.1016/j.asoc.2017.11.001}.
\newblock URL
  \url{http://linkinghub.elsevier.com/retrieve/pii/S1568494617306634}.

\bibitem[Mucha et~al.(2017)Mucha, Galaz, Mekyska, Kiska, Zvoncak, Smekal,
  Eliasova, Mrackova, Kostalova, Rektorova, Faundez-Zanuy, and
  Alonso-Hernandez]{mucha_identification_2017}
J.~Mucha, Z.~Galaz, J.~Mekyska, T.~Kiska, V.~Zvoncak, Z.~Smekal, I.~Eliasova,
  M.~Mrackova, M.~Kostalova, I.~Rektorova, M.~Faundez-Zanuy, and J.~B.
  Alonso-Hernandez.
\newblock Identification of hypokinetic dysarthria using acoustic analysis of
  poem recitation.
\newblock In \emph{2017 40th {International} {Conference} on
  {Telecommunications} and {Signal} {Processing} ({TSP})}, pages 739--742, July
  2017.
\newblock \doi{10.1109/TSP.2017.8076086}.

\bibitem[Nagrani et~al.(2017)Nagrani, Chung, and
  Zisserman]{nagrani_voxceleb:_2017}
A.~Nagrani, J.~S. Chung, and A.~Zisserman.
\newblock {VoxCeleb}: {A} {Large}-{Scale} {Speaker} {Identification} {Dataset}.
\newblock In \emph{Interspeech 2017}, pages 2616--2620. ISCA, Aug. 2017.
\newblock \doi{10.21437/Interspeech.2017-950}.
\newblock URL
  \url{http://www.isca-speech.org/archive/Interspeech_2017/abstracts/0950.html}.

\bibitem[Novotný et~al.(2014)Novotný, Rusz, Cmejla, and
  Ruzicka]{novotny_automatic_2014}
M.~Novotný, J.~Rusz, R.~Cmejla, and E.~Ruzicka.
\newblock Automatic {Evaluation} of {Articulatory} {Disorders} in {Parkinson}'s
  {Disease}.
\newblock \emph{IEEE/ACM Transactions on Audio, Speech, and Language
  Processing}, 22\penalty0 (9):\penalty0 1366--1378, Sept. 2014.
\newblock ISSN 2329-9290.
\newblock \doi{10.1109/TASLP.2014.2329734}.

\bibitem[Orozco-Arroyave et~al.(2014)Orozco-Arroyave, Hönig, Arias-Londoño,
  Bonilla, Skodda, Rusz, and Nöth]{orozco-arroyave_automatic_2014}
J.~R. Orozco-Arroyave, F.~Hönig, J.~D. Arias-Londoño, J.~F.~V. Bonilla,
  S.~Skodda, J.~Rusz, and E.~Nöth.
\newblock Automatic detection of {Parkinson}'s disease from words uttered in
  three different languages.
\newblock In \emph{{INTERSPEECH}}, pages 1573--1577, 2014.
\newblock URL
  \url{https://pdfs.semanticscholar.org/dbcb/d806177bfe6e09e05047e999422a1a0c79b3.pdf}.

\bibitem[Orozco-Arroyave et~al.(2015{\natexlab{a}})Orozco-Arroyave,
  Belalcazar-Bolaños, Arias-Londoño, Vargas-Bonilla, Skodda, Rusz, Daqrouq,
  Hönig, and Nöth]{orozco-arroyave_characterization_2015}
J.~R. Orozco-Arroyave, E.~A. Belalcazar-Bolaños, J.~D. Arias-Londoño, J.~F.
  Vargas-Bonilla, S.~Skodda, J.~Rusz, K.~Daqrouq, F.~Hönig, and E.~Nöth.
\newblock Characterization {Methods} for the {Detection} of {Multiple} {Voice}
  {Disorders}: {Neurological}, {Functional}, and {Laryngeal} {Diseases}.
\newblock \emph{IEEE Journal of Biomedical and Health Informatics}, 19\penalty0
  (6):\penalty0 1820--1828, Nov. 2015{\natexlab{a}}.
\newblock ISSN 2168-2194.
\newblock \doi{10.1109/JBHI.2015.2467375}.

\bibitem[Orozco-Arroyave et~al.(2015{\natexlab{b}})Orozco-Arroyave, Hönig,
  Arias-Londoño, Bonilla, Skodda, Rusz, and
  Nöth]{orozco-arroyave_voiced/unvoiced_2015}
J.~R. Orozco-Arroyave, F.~Hönig, J.~D. Arias-Londoño, J.~F.~V. Bonilla,
  S.~Skodda, J.~Rusz, and E.~Nöth.
\newblock Voiced/unvoiced transitions in speech as a potential bio-marker to
  detect parkinson's disease.
\newblock In \emph{{INTERSPEECH}}, pages 95--99. Citeseer, 2015{\natexlab{b}}.
\newblock URL
  \url{http://citeseerx.ist.psu.edu/viewdoc/download?doi=10.1.1.707.4297&rep=rep1&type=pdf}.

\bibitem[Orozco-Arroyave et~al.(2016)Orozco-Arroyave, Hönig, Arias-Londoño,
  Vargas-Bonilla, Daqrouq, Skodda, Rusz, and
  Nöth]{orozco-arroyave_automatic_2016}
J.~R. Orozco-Arroyave, F.~Hönig, J.~D. Arias-Londoño, J.~F. Vargas-Bonilla,
  K.~Daqrouq, S.~Skodda, J.~Rusz, and E.~Nöth.
\newblock Automatic detection of {Parkinson}'s disease in running speech spoken
  in three different languages.
\newblock \emph{The Journal of the Acoustical Society of America}, 139\penalty0
  (1):\penalty0 481--500, 2016.
\newblock URL \url{http://asa.scitation.org/doi/abs/10.1121/1.4939739}.

\bibitem[Povey et~al.(2011)Povey, Ghoshal, Boulianne, Burget, Glembek, Goel,
  Hannemann, Motlicek, Qian, Schwarz, Silovsky, Stemmer, and
  Vesely]{povey_kaldi_2011}
D.~Povey, A.~Ghoshal, G.~Boulianne, L.~Burget, O.~Glembek, N.~Goel,
  M.~Hannemann, P.~Motlicek, Y.~Qian, P.~Schwarz, J.~Silovsky, G.~Stemmer, and
  K.~Vesely.
\newblock The {Kaldi} {Speech} {Recognition} {Toolkit}.
\newblock In \emph{{IEEE} 2011 {Workshop} on {Automatic} {Speech} {Recognition}
  and {Understanding}}, page~4, 2011.

\bibitem[Prince(2007)]{prince_probabilistic_2007}
S.~J.~D. Prince.
\newblock Probabilistic {Linear} {Discriminant} {Analysis} for.
\newblock In \emph{Inferences {About} {Identity} ,” {ICCV}}, 2007.

\bibitem[Quatieri(2001)]{quatieri_discrete-time_2001}
T.~F. Quatieri.
\newblock \emph{Discrete-{Time} {Speech} {Signal} {Processing}: {Principles}
  and {Practice}}.
\newblock Prentice Hall, Upper Saddle River, NJ, 1 edition edition, Nov. 2001.
\newblock ISBN 978-0-13-242942-9.

\bibitem[Reynolds et~al.(2000)Reynolds, Quatieri, and
  Dunn]{reynolds_speaker_2000}
D.~A. Reynolds, T.~F. Quatieri, and R.~B. Dunn.
\newblock Speaker {Verification} {Using} {Adapted} {Gaussian} {Mixture}
  {Models}.
\newblock \emph{Digital Signal Processing}, 10\penalty0 (1-3):\penalty0 19--41,
  Jan. 2000.
\newblock ISSN 10512004.
\newblock \doi{10.1006/dspr.1999.0361}.
\newblock URL
  \url{http://linkinghub.elsevier.com/retrieve/pii/S1051200499903615}.

\bibitem[Rizvi et~al.(2020)Rizvi, Nissar, Masood, Ahmed, and
  Ahmad]{rizvi_lstm_2020}
D.~R. Rizvi, I.~Nissar, S.~Masood, M.~Ahmed, and F.~Ahmad.
\newblock An {LSTM} based {Deep} learning model for voice-based detection of
  {Parkinson}’s disease.
\newblock \emph{International Journal of Advanced Science and Technology},
  29\penalty0 (5):\penalty0 8, 2020.

\bibitem[Rusz et~al.(2011)Rusz, Čmejla, Růžičková, Klempíř, Majerová,
  Picmausová, Roth, and Růžička]{rusz_acoustic_2011}
J.~Rusz, R.~Čmejla, H.~Růžičková, J.~Klempíř, V.~Majerová,
  J.~Picmausová, J.~Roth, and E.~Růžička.
\newblock Acoustic assessment of voice and speech disorders in {Parkinson}'s
  disease through quick vocal test.
\newblock \emph{Movement Disorders}, 26\penalty0 (10):\penalty0 1951--1952,
  Aug. 2011.
\newblock ISSN 08853185.
\newblock \doi{10.1002/mds.23680}.
\newblock URL \url{http://doi.wiley.com/10.1002/mds.23680}.

\bibitem[Rusz et~al.(2013)Rusz, Cmejla, Tykalova, Ruzickova, Klempir, Majerova,
  Picmausova, Roth, and Ruzicka]{rusz_imprecise_2013}
J.~Rusz, R.~Cmejla, T.~Tykalova, H.~Ruzickova, J.~Klempir, V.~Majerova,
  J.~Picmausova, J.~Roth, and E.~Ruzicka.
\newblock Imprecise vowel articulation as a potential early marker of
  {Parkinson}'s disease: {Effect} of speaking task.
\newblock \emph{The Journal of the Acoustical Society of America}, 134\penalty0
  (3):\penalty0 2171--2181, Sept. 2013.
\newblock ISSN 0001-4966.
\newblock \doi{10.1121/1.4816541}.
\newblock URL
  \url{http://scitation.aip.org/content/asa/journal/jasa/134/3/10.1121/1.4816541}.

\bibitem[Rusz et~al.(2015{\natexlab{a}})Rusz, Bonnet, Klempíř, Tykalová,
  Baborová, Novotný, Rulseh, and Růžička]{rusz_speech_2015}
J.~Rusz, C.~Bonnet, J.~Klempíř, T.~Tykalová, E.~Baborová, M.~Novotný,
  A.~Rulseh, and E.~Růžička.
\newblock Speech disorders reflect differing pathophysiology in {Parkinson}’s
  disease, progressive supranuclear palsy and multiple system atrophy.
\newblock \emph{Journal of Neurology}, 262\penalty0 (4):\penalty0 992--1001,
  Apr. 2015{\natexlab{a}}.
\newblock ISSN 0340-5354, 1432-1459.
\newblock \doi{10.1007/s00415-015-7671-1}.
\newblock URL \url{http://link.springer.com/10.1007/s00415-015-7671-1}.

\bibitem[Rusz et~al.(2015{\natexlab{b}})Rusz, Hlavnička, Tykalová, Bušková,
  Ulmanová, Růžička, and Šonka]{rusz_quantitative_2015}
J.~Rusz, J.~Hlavnička, T.~Tykalová, J.~Bušková, O.~Ulmanová,
  E.~Růžička, and K.~Šonka.
\newblock Quantitative assessment of motor speech abnormalities in idiopathic
  rapid eye movement sleep behaviour disorder.
\newblock \emph{Sleep Medicine}, Sept. 2015{\natexlab{b}}.
\newblock ISSN 13899457.
\newblock \doi{10.1016/j.sleep.2015.07.030}.
\newblock URL
  \url{http://linkinghub.elsevier.com/retrieve/pii/S1389945715009296}.

\bibitem[Sakar et~al.(2013)Sakar, Isenkul, Sakar, Sertbas, Gurgen, Delil,
  Apaydin, and Kursun]{sakar_collection_2013}
B.~Sakar, M.~Isenkul, C.~Sakar, A.~Sertbas, F.~Gurgen, S.~Delil, H.~Apaydin,
  and O.~Kursun.
\newblock Collection and {Analysis} of a {Parkinson} {Speech} {Dataset} {With}
  {Multiple} {Types} of {Sound} {Recordings}.
\newblock \emph{IEEE Journal of Biomedical and Health Informatics}, 17\penalty0
  (4):\penalty0 828--834, July 2013.
\newblock ISSN 2168-2194.
\newblock \doi{10.1109/JBHI.2013.2245674}.

\bibitem[Sakar et~al.(2017)Sakar, Serbes, and Sakar]{sakar_analyzing_2017}
B.~E. Sakar, G.~Serbes, and C.~O. Sakar.
\newblock Analyzing the effectiveness of vocal features in early telediagnosis
  of {Parkinson}'s disease.
\newblock \emph{PLOS ONE}, 12\penalty0 (8):\penalty0 e0182428, Aug. 2017.
\newblock ISSN 1932-6203.
\newblock \doi{10.1371/journal.pone.0182428}.
\newblock URL
  \url{https://journals.plos.org/plosone/article?id=10.1371/journal.pone.0182428}.

\bibitem[Snyder et~al.(2016)Snyder, Ghahremani, Povey, Garcia-Romero, Carmiel,
  and Khudanpur]{snyder_deep_2016}
D.~Snyder, P.~Ghahremani, D.~Povey, D.~Garcia-Romero, Y.~Carmiel, and
  S.~Khudanpur.
\newblock Deep neural network-based speaker embeddings for end-to-end speaker
  verification.
\newblock In \emph{2016 {IEEE} {Spoken} {Language} {Technology} {Workshop}
  ({SLT})}, pages 165--170, San Diego, CA, Dec. 2016. IEEE.
\newblock ISBN 978-1-5090-4903-5.
\newblock \doi{10.1109/SLT.2016.7846260}.
\newblock URL \url{http://ieeexplore.ieee.org/document/7846260/}.

\bibitem[Snyder et~al.(2017)Snyder, Garcia-Romero, Povey, and
  Khudanpur]{snyder_deep_2017}
D.~Snyder, D.~Garcia-Romero, D.~Povey, and S.~Khudanpur.
\newblock Deep {Neural} {Network} {Embeddings} for {Text}-{Independent}
  {Speaker} {Verification}.
\newblock In \emph{Interspeech 2017}, pages 999--1003. ISCA, Aug. 2017.
\newblock \doi{10.21437/Interspeech.2017-620}.
\newblock URL
  \url{http://www.isca-speech.org/archive/Interspeech_2017/abstracts/0620.html}.

\bibitem[Snyder et~al.(2018{\natexlab{a}})Snyder, Garcia-Romero, McCree, Sell,
  Povey, and Khudanpur]{snyder_spoken_2018}
D.~Snyder, D.~Garcia-Romero, A.~McCree, G.~Sell, D.~Povey, and S.~Khudanpur.
\newblock Spoken {Language} {Recognition} using {X}-vectors.
\newblock In \emph{Odyssey 2018 {The} {Speaker} and {Language} {Recognition}
  {Workshop}}, pages 105--111. ISCA, June 2018{\natexlab{a}}.
\newblock \doi{10.21437/Odyssey.2018-15}.
\newblock URL
  \url{http://www.isca-speech.org/archive/Odyssey_2018/abstracts/38.html}.

\bibitem[Snyder et~al.(2018{\natexlab{b}})Snyder, Garcia-Romero, Sell, Povey,
  and Khudanpur]{snyder_x-vectors:_2018}
D.~Snyder, D.~Garcia-Romero, G.~Sell, D.~Povey, and S.~Khudanpur.
\newblock X-{Vectors}: {Robust} {DNN} {Embeddings} for {Speaker} {Recognition}.
\newblock In \emph{2018 {IEEE} {International} {Conference} on {Acoustics},
  {Speech} and {Signal} {Processing} ({ICASSP})}, pages 5329--5333, Calgary,
  AB, Apr. 2018{\natexlab{b}}. IEEE.
\newblock ISBN 978-1-5386-4658-8.
\newblock \doi{10.1109/ICASSP.2018.8461375}.
\newblock URL \url{https://ieeexplore.ieee.org/document/8461375/}.

\bibitem[Tremblay et~al.(2019)Tremblay, Abbasi, Zeighami, and
  Dagher]{tremblay_gender_2019}
C.~Tremblay, N.~Abbasi, Y.~Zeighami, and A.~Dagher.
\newblock gender difference in brain atrophy in de novo parkinson's
  disease.pdf.
\newblock Italy, Rome, 2019.
\newblock URL
  \url{https://ww5.aievolution.com/hbm1901/index.cfm?do=abs.viewAbs&abs=3198}.

\bibitem[Tsanas et~al.(2011)Tsanas, Little, McSharry, and
  Ramig]{tsanas_nonlinear_2011}
A.~Tsanas, M.~A. Little, P.~E. McSharry, and L.~O. Ramig.
\newblock Nonlinear speech analysis algorithms mapped to a standard metric
  achieve clinically useful quantification of average {Parkinson}'s disease
  symptom severity.
\newblock \emph{Journal of The Royal Society Interface}, 8\penalty0
  (59):\penalty0 842--855, June 2011.
\newblock ISSN 1742-5689, 1742-5662.
\newblock \doi{10.1098/rsif.2010.0456}.
\newblock URL
  \url{http://rsif.royalsocietypublishing.org/cgi/doi/10.1098/rsif.2010.0456}.

\bibitem[Tsanas et~al.(2012)Tsanas, Little, McSharry, Spielman, and
  Ramig]{tsanas_novel_2012}
A.~Tsanas, M.~A. Little, P.~E. McSharry, J.~Spielman, and L.~O. Ramig.
\newblock Novel {Speech} {Signal} {Processing} {Algorithms} for
  {High}-{Accuracy} {Classification} of {Parkinson}'s {Disease}.
\newblock \emph{IEEE Transactions on Biomedical Engineering}, 59\penalty0
  (5):\penalty0 1264--1271, May 2012.
\newblock ISSN 0018-9294, 1558-2531.
\newblock \doi{10.1109/TBME.2012.2183367}.
\newblock URL
  \url{http://ieeexplore.ieee.org/lpdocs/epic03/wrapper.htm?arnumber=6126094}.

\bibitem[Vásquez-Correa et~al.(2017)Vásquez-Correa, Orozco-Arroyave, and
  Nöth]{vasquez-correa_convolutional_2017}
J.~Vásquez-Correa, J.~R. Orozco-Arroyave, and E.~Nöth.
\newblock Convolutional {Neural} {Network} to {Model} {Articulation}
  {Impairments} in {Patients} with {Parkinson}’s {Disease}.
\newblock In \emph{{INTERSPEECH}}, pages 314--318. ISCA, Aug. 2017.
\newblock \doi{10.21437/Interspeech.2017-1078}.
\newblock URL
  \url{http://www.isca-speech.org/archive/Interspeech_2017/abstracts/1078.html}.

\bibitem[Zhang et~al.(2018)Zhang, Wang, Li, and Xu]{zhang_deepvoice:_2018}
H.~Zhang, A.~Wang, D.~Li, and W.~Xu.
\newblock {DeepVoice}: {A} voiceprint-based mobile health framework for
  {Parkinson}'s disease identification.
\newblock In \emph{2018 {IEEE} {EMBS} {International} {Conference} on
  {Biomedical} \& {Health} {Informatics} ({BHI})}, pages 214--217, Las Vegas,
  NV, USA, Mar. 2018. IEEE.
\newblock ISBN 978-1-5386-2405-0.
\newblock \doi{10.1109/BHI.2018.8333407}.
\newblock URL \url{http://ieeexplore.ieee.org/document/8333407/}.

\end{thebibliography}

\end{document}